\documentclass[conference]{IEEEtran}
\IEEEoverridecommandlockouts
\usepackage{cite}
\usepackage{amsmath,amssymb,amsfonts}
\usepackage{algorithmic}
\usepackage{graphicx}
\usepackage{textcomp}
\usepackage{xcolor}
\usepackage{xspace}
\usepackage{siunitx}
\usepackage{array}
\usepackage{stfloats}
\usepackage{url}
\usepackage{verbatim}
\usepackage{bm}
\usepackage{mathrsfs}
\usepackage{booktabs}
\usepackage{svg}
\usepackage[ruled,lined]{algorithm2e}
\usepackage{makecell}
\usepackage{multirow}
\usepackage{arydshln}
\usepackage{stackengine}
\usepackage[export]{adjustbox}
\usepackage{subcaption}

\usepackage[pagebackref,breaklinks]{hyperref}
\usepackage[capitalize]{cleveref}

\makeatletter
\DeclareRobustCommand\onedot{\futurelet\@let@token\@onedot}
\def\@onedot{\ifx\@let@token.\else.\null\fi\xspace}

\def\eg{\emph{e.g}\onedot} 
\def\ie{\emph{i.e}\onedot}

\def\etal{\emph{et al}\onedot}
\makeatother

\def\BibTeX{{\rm B\kern-.05em{\sc i\kern-.025em b}\kern-.08em
    T\kern-.1667em\lower.7ex\hbox{E}\kern-.125emX}}
\begin{document}

\title{Encryption and Authentication with a Lensless Camera Based on a Programmable Mask}

\author{Eric Bezzam and Martin Vetterli\\
\textit{Audiovisual Communications Laboratory}\\
\textit{École Polytechnique Fédérale de Lausanne (EPFL)}\\
Lausanne, Switzerland\\
\texttt{first.last@epfl.ch}}

\maketitle

\begin{abstract}
Lensless cameras replace traditional optics with thin masks, leading to highly multiplexed measurements akin to encryption. However, static masks in conventional designs leave systems vulnerable to simple attacks. This work explores the use of programmable masks to enhance security by dynamically varying the mask patterns. We perform our experiments with a low-cost system (around 100 USD) based on a liquid crystal display. Experimental results demonstrate that variable masks successfully block a variety of attacks while enabling high-quality recovery for legitimate users. The system’s encryption strength exceeds AES-256, achieving effective key lengths over 2'500 bits. Additionally, we demonstrate how a programmable mask enables robust authentication and verification, as each mask pattern leaves a unique fingerprint on the image. When combined with a lensed system, lensless measurements can serve as analog certificates, providing a novel solution for verifying image authenticity and combating deepfakes.
\end{abstract}


\section{Introduction}
\label{sec:intro}

Cameras are everywhere, and are increasingly integrated with cloud-based systems for analysis and processing. 
This connectivity enhances convenience, but also heightens the risk of data hacks and leaks.
While conventional encryption provides strong digital security, 
we investigate the use of analog hardware as a first layer of security.
Using optics has garnered interest due to the many degrees of freedom that make it harder to attack and the ``free'' compute from light propagation~\cite{Javidi_2016}.
Lensless cameras, which shift image formation from optics to digital post-processing, offer one such solution~\cite{boominathan2022recent}.
By replacing traditional optics with thin masks,
their multiplexed measurements are inherently indecipherable, making them attractive for privacy and security applications~\cite{Tan2020,10666814}.
However, existing designs 
rely on static masks, leaving systems susceptible to attacks that estimate the point spread function (PSF) to decode measurements,
or that attempt to learn decoding from many measurements.

The strength of our work is in its simplicity.
While state-of-the-art lensless imaging systems rely on static, uniquely fabricated masks for high-resolution recovery~\cite{phlatcam,Lee:23}, 
programmable masks are underexplored. 
With such a mask, the system can be easily reconfigured to counteract potential leaks and thwart adversarial attacks. 
Programmable masks can also enable novel applications in image authentication. 
Each mask leaves a unique ``fingerprint'' on the captured image, 
which can be used to verify its origin or timestamp. 
Unlike lensed systems, for which authentication relies on detecting subtle hardware artifacts~\cite{1634362,10.1117/12.649775,6126225}, lensless cameras intentionally introduce distortions which can be more readily detected. 
With consumer devices increasingly integrating complementary sensors, such for infrared, LiDAR, and ambient light, the application of lensless cameras as proposed here has the potential to serve 
as a robust tool for verifying image authenticity, 
addressing critical challenges like deepfake proliferation.
Our contributions include:

\begin{figure}[t!]
	\centering
	\includegraphics[width=\linewidth]{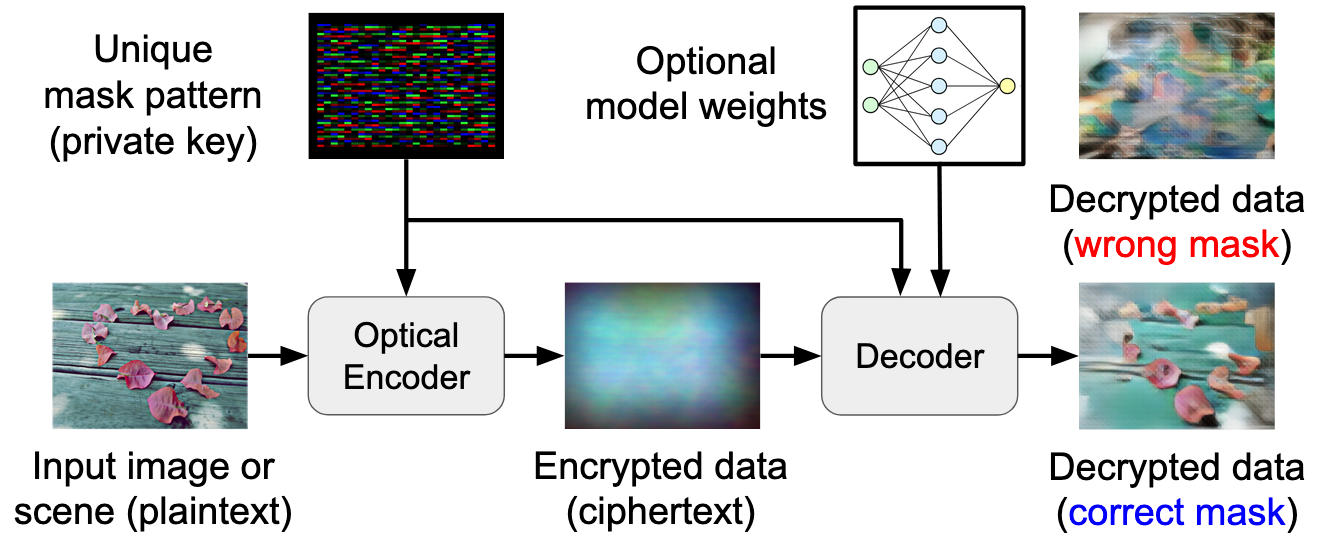}
	\caption{Example encryption and decryption with a lensless camera based on a programmable mask. The optical encoder captures an encrypted image with a unique mask pattern. For decryption, the corresponding mask is needed by the decoder, which can optionally use learning-based algorithms for improved imaging/decryption quality. 
	}
	\label{fig:overview}
\end{figure}

\begin{enumerate}
	\item Theoretical and empirical analysis of the encryption strength of a lensless system with a programmable mask, surpassing AES-256 with effective key lengths exceeding 2'500 bits for our prototype system.
	\item Empirical validation of the system's robustness against diverse attacks, including chosen-plaintext, known-plaintext, ciphertext-only, and brute-force methods.
	\item A novel application for lensless imaging: robust authentication enabled by programmable masks, leveraging the unique ``fingerprints'' they impart on captured images.
\end{enumerate}
We validate our approach with \textit{DigiCam}~\cite{Bezzam2024}, a low-cost prototype (around 100 USD) of a programmable mask-based system.
Using such an affordable system demonstrates the viability and potential scalability of such a solution.

\section{Related Work}
\label{sec:related}

\subsection{Optical Encryption}

Encryption transforms information into a form accessible only to authorized entities.
While traditionally implemented digitally, analog optical methods are also effective,
such as downsampling~\cite{Ryoo2017},
introducing aberrations~\cite{Hinojosa2022},
or using optical masks~\cite{Refregier:95,9025454}.
In cryptographic terms, the image to be encoded is the \textit{plaintext},
and the encrypted output is the \textit{ciphertext}.
When using optical masks, the pattern or its PSF is a form of \textit{private key}, 
which may be needed for encryption and decryption as in symmetric-key algorithms.
An example encryption and decryption pipeline is visualized in \cref{fig:overview}.

Many optical encryption solutions, such as double random phase encoding (DRPE)~\cite{Refregier:95}, require coherent illumination and precise alignment, limiting their practicality in real-world settings.  
Khan \etal~\cite{10666814} proposed a lensless system for optical encryption under incoherent illumination. 
Various attacks can be performed to determine a system's encryption strength: 
(1) chosen-plaintext attacks (CPAs) where an attack can choose arbitrary plaintexts to be encrypted,
(2) known-plaintext attacks (KPAs) where the attacker cannot choose the plaintexts but has pairs of plaintexts and ciphertexts,
(3) ciphertext-only attacks (COAs) where the attacker only has access to encrypted data,
and (4) brute force attacks that try determining the secret key or decode an image by trial-and-error.

\subsection{Lensless Imaging}
\label{sec:lensless_background}

Lensless cameras shift image formation from the analog optics to the digital post-processing.
Imaging typically involves solving a deconvolution problem  expressed as:
\begin{align}
	\label{eq:opt_gen}
	\bm{\hat{X}} = \arg \min_{\bm{X}} \frac{1}{2} ||\bm{Y} - \bm{P} \ast \bm{X}||_2^2 + \lambda \mathcal{R}(\bm{X}),
\end{align}
where $\bm{Y}$ is the lensless measurement,
$\bm{P}$ is the system's PSF,  
$\ast$ represents 2D convolution,
and $\mathcal{R}(\cdot)$ is a regularization function.
\cref{eq:opt_gen} can be solved in closed-form with $\ell_2$-regularization~\cite{flatcam} or Wiener filtering. However, for improved quality, iterative methods like 
ADMM~\cite{ADMM} are employed, incorporating advanced regularizers such as total variation minimization~\cite{Antipa:18}.
More recent approaches combine physical models with deep learning, achieving superior results while reducing imaging time~\cite{Monakhova:19,9239993,Li:23}.

\subsection{Optical Watermarking and Camera Identification}

Optical encryption techniques, \eg based on DRPE~\cite{Kishk:02},
have also been applied to watermarking, enabling the embedding of secret information for verifying image authenticity. 
However, these approaches often require coherent illumination and precise alignment, limiting their practicality.
An alternative for image authenticity verification is image forensics, which identifies camera sources by analyzing artifacts introduced by the imaging system, such as fixed pattern noise~\cite{1634362}, aberrations~\cite{10.1117/12.649775} or post-processing artifacts~\cite{6126225}.
These approaches typically focus on subtle artifacts to avoid affecting the primary content.
Lensless imaging, on the other hand, reconstructs images from significantly distorted measurements, allowing the deliberate inclusion of more prominent, easily detectable fingerprints, as investigated in this work. 

\section{Methodology}
\label{sec:method}
In \cref{sec:programmable}, we introduce generalizable lensless imaging with a programmable mask, which forms the basis of our encryption and decryption methods. \cref{sec:method_encryption_strength,sec:method_auth} discuss encryption and authentication respectively.

\subsection{Generalizable Lensless Imaging with a Programmable Mask as an Encryption Scheme}
\label{sec:programmable}

\begin{figure}[t!]
	\centering
	\includegraphics[width=\linewidth]{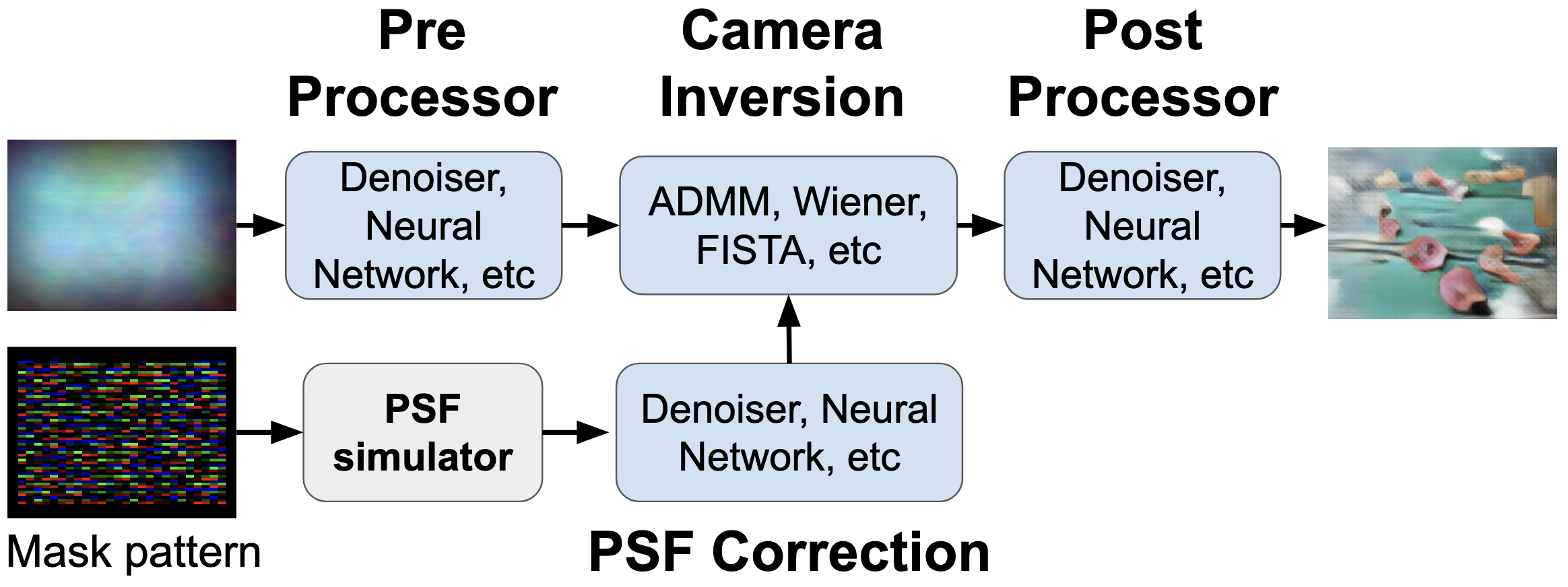}
	\caption{Modular lensless imaging pipeline as proposed in~\cite{Bezzam2024}, which can serve as the decoder in \cref{fig:overview}. Blocks in blue are learned from measured data.}
	\label{fig:modular}
\end{figure}

A significant limitation of previous designs is their reliance on static optical configurations~\cite{Tan2020,10666814}. While they can be resistant to brute-force attacks due to a large key space for potential masks, they remain vulnerable to targeted attacks or data leaks. We address these vulnerabilities by introducing a programmable mask, enabling dynamic modifications to the optical configuration. This approach strengthens security and mitigates risks from potential system parameter or data leaks.

To ensure compatibility with changing masks, the decoder must handle variations in the PSF.
Traditional optimization-based methods, 
such as ADMM, can solve \cref{eq:opt_gen} for dynamic masks but yield subpar image quality. 
In~\cite{Bezzam2024}, the authors propose a modular learned reconstruction framework, as visualized in \cref{fig:modular}, 
which accommodates variable mask patterns. 
This framework requires two key components:
(1) a realistic simulator to generate the imaging system’s PSF given a mask pattern,
and (2) a dataset of tuples comprising lensed images, lensless measurements, and corresponding mask patterns (\ie plaintext-ciphertext-key) to train the decoder.

Using only the mask pattern for encoding and decoding is similar to symmetric-key encryption, where both encryption (imaging system) and decryption (user) require the same private key. 
Introducing a learned model extends this analogy to a public-key system, 
where the model parameters can act as a \textit{public key} shared across users for a particular camera or application, as shown in \cref{fig:overview}.

\subsection{Encryption Strength of a Programmable Mask System}
\label{sec:method_encryption_strength}

To evaluate the encryption strength of a programmable mask system, we consider an adversary attempting to brute-force the PSF for decryption. 
Using insights from model mismatch analysis~\cite{9546648}, we examine the consequences of decoding with an incorrect PSF. The imaging process can be modeled, in the most general sense (without the convolution simplification as in \cref{eq:opt_gen}), 
as a matrix-vector product with additive noise:
\begin{align}
    \bm{y} = \bm{Hx} + \bm{n},
\end{align}
where $\bm{H}$ represents the system matrix.
If decoding is performed with a perturbed matrix $\hat{\bm{H}}=(\bm{H}-\bm{\Delta})$, applying direct inversion yields:
\begin{align}
 \label{eq:psf_err}
	\bm{\hat{x}} = \hat{\bm{H}}^{-1}\bm{y}
	= \bm{x} + \bm{H}^{-1}\bm{n} + \underbrace{( \bm{x} + \bm{H}^{-1}\bm{n})\sum_{k=1}^{\infty}(\bm{H}^{-1}\bm{\Delta})^k}_{\text{error when using wrong system matrix}}.
\end{align}
A detailed derivation is provided \cref{app:deviation}.
As expected, larger deviations $\bm{\Delta}$ results in a large perturbation.
This breakdown extends to other inversion methods, 
including Wiener filtering, gradient descent, and ADMM~\cite{Bezzam2024}.

Unlike standard encryption schemes like the Advanced Encryption Standard (AES), 
where a single bit error in the key renders meaningless outputs,
lensless imaging with a slightly perturbed PSF (\ie small $\bm{\Delta}$) can still produce a recognizable image. 
Therefore, it is not sufficient to count the degrees of freedom for determining the strength of an optical encryption scheme. Rather the deviation $\bm{\Delta}$ must be quantified, \ie the deviation for which the recovered image with,  \eg \cref{eq:psf_err} or with a perturbed PSF in \cref{eq:opt_gen}, 
becomes indecipherable.

For a programmable mask with $N$ pixels of bit-depth $b$, there are $b^N$ possible mask patterns, \ie degrees of freedom.
However, correctly determining all $N$ pixels may not be necessary for a discernible image,
\ie for a brute-force attack.
If $W \in (0, 1]$ represents the minimum ratio of correctly determined pixels for a discernible image,
to obtain an encryption strength better or equivalent to a key size of $K$: $b^{NW} \geq  2^K$.
Isolating for $W$ gives a lower bound on the ratio needed for a search space equivalent to a key of size $K$:
\begin{align}
    \label{eq:correct_bound}
	W &\geq \frac{K\log_b2}{N}.
\end{align}

\subsection{Robust Authentication with a Lensless Certificate}
\label{sec:method_auth}
Programmable masks can be used as a form of optical fingerprinting with unique mask patterns during measurement.
Similar to two-factor authentication, 
a mask pattern specific to a user/camera and timestamp can be generated:
\begin{equation}
	\bm{W}_{u,t} = \text{generate}(\text{user}, \text{timestamp}),
\end{equation}
where $\bm{W}_{u,t}$ are mask pattern values that can be set to the camera's programmable mask.
For a given lensless measurement $\bm{Y}$, 
we compute a score to determine if it was captured by a specific user $u'$ and at timestamp $t'$,
using a candidate mask pattern $\bm{W}_{u',t'} $.
A simple authentication metric can be based on data fidelity:
\begin{equation}
	\label{eq:auth}
	\text{auth}(\bm{Y}, \bm{W}_{u',t'}, \bm{\theta}) = ||\bm{Y} - \bm{P}_{u',t'} \ast \text{dec}(\bm{Y}, \bm{W}_{u',t'}, \bm{\theta})||_2^2,
\end{equation}
where $\bm{\theta}$ are the model/recovery parameters,  $\bm{P}_{u',t'}$ is the PSF corresponding to the mask pattern $\bm{W}_{u',t'}$, and $\text{dec}(\cdot)$ decodes the ciphertext $\bm{Y}$, \ie the output of image recovery.
The mask pattern provided by the true owner/capturer of the image $\bm{Y}$ minimizes \cref{eq:auth},
while a false mask pattern will lead to a larger score.
Intuitively, this can be linked back to  \cref{eq:psf_err} where error in the decoded output is increased when model deviation is increased.

If the lensless camera is deployed alongside a lensed camera,
additional metrics can be computed by comparing the lensless reconstruction to the lensed measurement:
\begin{align}
    \label{eq:auth_ref}
    \text{auth}(\bm{Y}, \bm{W}_{u',t'}, \bm{\theta}, \bm{X}_{\text{lensed}}) = \mathcal{L}(\bm{X}_{\text{lensed}}, \text{dec}(\bm{Y}, \bm{W}_{u',t'}, \bm{\theta})),
\end{align}
where $\mathcal{L}$ is a similarity metric, \eg mean-squared error (MSE), 
structural similarity index measure (SSIM),
or learned perceptual image patch similarity (LPIPS) 
which computes similarity with pre-trained VGG networks as features extractors and lower is better within $[0,1]$~\cite{zhang2018perceptual}.
When deployed with a lensed camera, the lensless image can serve as a ``certificate,'' authenticating the paired lensed image.
This unique functionality offers a solution to challenges such as deepfake validation.

\section{Experiments and Results}
\label{sec:results}

We conduct the following experiments using the \textit{DigiCam} system~\cite{Bezzam2024}, a lensless camera prototype that uses a low-cost liquid crystal display (LCD) as a programmable mask:
\begin{enumerate}
	\item \textbf{Variable Mask Patterns} (\cref{sec:variable_exp}): We assess the system’s resilience against known-plaintext attacks (KPAs) and chosen-plaintext attack (CPAs) by demonstrating how varying mask patterns protect against PSF estimation and decoder training.
	\item \textbf{Encryption Strength} (\cref{sec:encryption_exp}): We vary the mask pattern error to analyze the encryption strength, as discussed in \cref{sec:method_encryption_strength},
    and show robustness to brute-force ciphertext-only attacks (COAs).
	\item \textbf{Authentication} (\cref{sec:auth_exp}): We validate the authentication framework described in \cref{sec:method_auth}, showcasing the ability to authenticate images based on unique mask pattern fingerprints.
\end{enumerate}

\subsection{Experimental Setup}

\subsubsection{Datasets}

Our experiments use the \textit{MirFlickr-S} and \textit{MirFlickr-M} datasets, both collected using the \textit{DigiCam} system~\cite{Bezzam2024}. 
These consist of 25'000 images measured from the \textit{MirFlickr} dataset~\cite{huiskes2008mir}, 
displayed on a screen placed \SI{30}{\centi\meter} from the camera. 
Both datasets share an \SI{85}{\percent}-\SI{15}{\percent} train-test split (21'250-3'750 files) of identical content but differ in mask patterns during measurement:
\textit{MirFlickr-S} uses the same mask pattern for both training and test measurement,
while \textit{MirFlickr-M} uses $85$ unique mask patterns for the training set and another $15$ distinct mask patterns for the test set (250 measurements per mask).
All patterns are generated according to a uniform distribution.
During image recovery, 
PSFs corresponding to the mask patterns are simulated using the method outlined in~\cite{Bezzam2024}. Example simulated PSFs are shown in \cref{fig:exp2_encrypt} (top row).

\subsubsection{Image Recovery Methods}

For the decoder in \cref{fig:overview},
we consider: 
(1) 100 iterations of ADMM with fixed hyperparameters (no training required),
and (2) the modular learned approach shown in \cref{fig:modular}.
We refer to the two models as \textit{ADMM} and \textit{Learned}.
The latter uses a denoising residual U-Net (DRUNet) architecture~\cite{zhang2021plug} for PSF correction (128K parameters) and for the pre- and post-processors (3.9M and 4.1M parameters respectively), and five unrolled iterations of ADMM for camera inversion~\cite{Monakhova:19},
for a total of around 8.1M parameters.
PyTorch is used for training and evaluation with an Intel Xeon E5-2680 v3 \SI{2.5}{\giga\hertz} CPU
and 4$\times$ Nvidia Titan X Pascal GPUs. 
The Adam optimizer is used with a learning rate of $10^{-4}$, $\beta_1=0.9$, $\beta_2=0.999$ for $25$ epochs and a batch size of $4$.
As evaluation metrics, we use peak signal-to-noise ratio (PSNR) in decibels,
SSIM, and LPIPS.

\subsection{Variable Mask to Defend Against Data Leaks}
\label{sec:variable_exp}

\begin{figure}[t!]
\centering
    \newcommand{\figsizeunet}{0.18}
    \newcommand{\newlineunet}{10pt}
	\renewcommand{\arraystretch}{1} 
	\setlength{\tabcolsep}{0.1em} 
	\begin{tabular}{ccccc}
    &   
    \multicolumn{2}{c}{\textbf{Correct PSF}}
    & \multicolumn{2}{c}{\textbf{Wrong PSF}}
    \\ 
    \cmidrule(r){2-3} \cmidrule(r){4-5} 
    \makecell{\textbf{Original}\\\textbf{+Lensless}}  & \textit{ADMM} & \textit{Learned} &\textit{ADMM} & \textit{Learned} \\
\includegraphics[width=\figsizeunet\linewidth,valign=m]{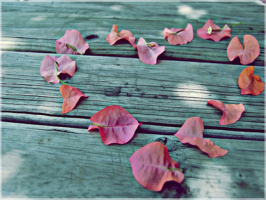}
\llap{\shortstack[l]{%
	\hspace{-1.2cm}\includegraphics[scale=.03]{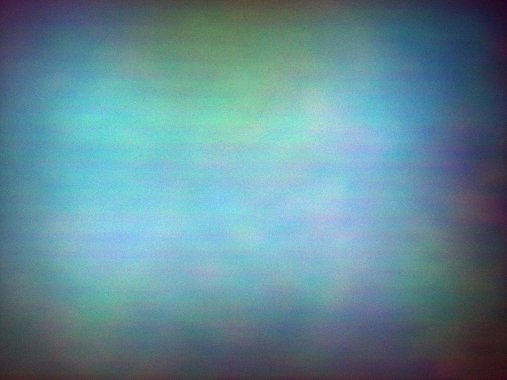}\\
		\rule{0ex}{0.13in}%
	}
	\rule{0.2in}{0ex}}

& 
    \includegraphics[width=\figsizeunet\linewidth,valign=m]{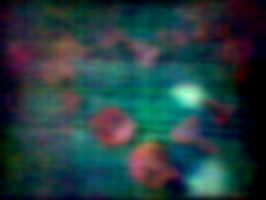}& 
    \includegraphics[width=\figsizeunet\linewidth,valign=m]{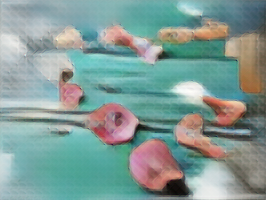}& 
    \includegraphics[width=\figsizeunet\linewidth,valign=m]{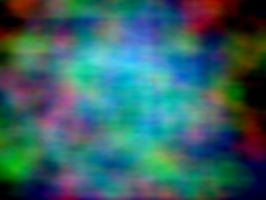}
    & 
    \includegraphics[width=\figsizeunet\linewidth,valign=m]{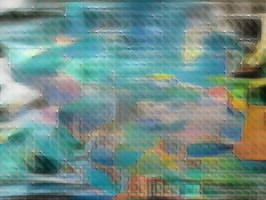}
    \\[\newlineunet]
    \includegraphics[width=\figsizeunet\linewidth,valign=m]{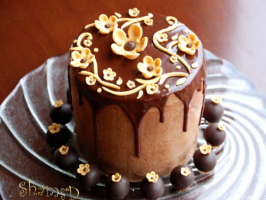}
\llap{\shortstack[l]{%
	\hspace{-1.2cm}\includegraphics[scale=.03]{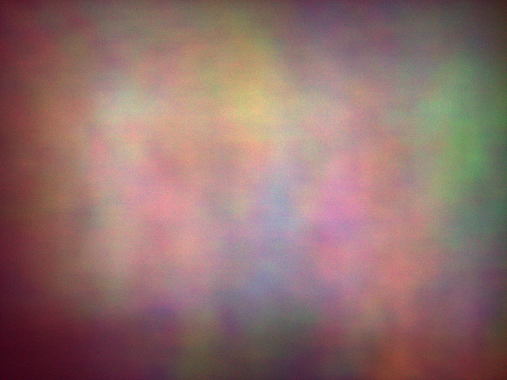}\\
		\rule{0ex}{0.13in}%
	}
	\rule{0.2in}{0ex}}
    & 
    \includegraphics[width=\figsizeunet\linewidth,valign=m]{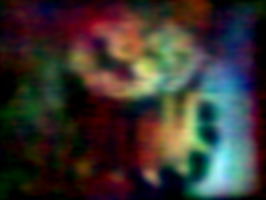}& 
    \includegraphics[width=\figsizeunet\linewidth,valign=m]{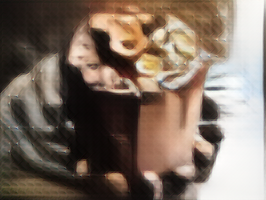}
    & 
    \includegraphics[width=\figsizeunet\linewidth,valign=m]{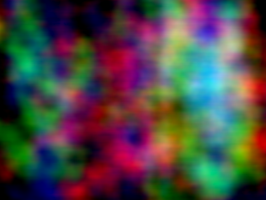}
    & 
    \includegraphics[width=\figsizeunet\linewidth,valign=m]{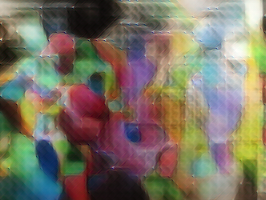}
    \\
	\end{tabular}
	\caption{Example reconstructions with correct and wrong PSFs.}
    \label{fig:exp1_admm}
\end{figure}

The visual privacy of lensless imaging can be observed in the raw lensless measurements shown in \cref{fig:exp1_admm} (insets of left-most column), which are indiscernible due to the multiplexing characteristic of such cameras.
However, this privacy can be compromised if an adversary obtains the system's PSF, either through CPAs~\cite{10666814} or because of a leak.
With access to a PSF estimate, adversaries can reconstruct discernible images using solvers like ADMM, 
as demonstrated in \cref{fig:exp1_admm} (second column).
Howoever, reconstruction quality degrades significantly when using an incorrect PSF (fourth column of \cref{fig:exp1_admm}).
In the event of a PSF leak, a programmable system like \textit{DigiCam} offers an added layer of security, allowing mask/PSF reconfiguration effectively render the leaked PSF obsolete.

\begin{figure}[t!]
\centering
    \newcommand{\figsizeunet}{0.23}
    \newcommand{\newlineunet}{15pt}
	\renewcommand{\arraystretch}{1} 
	\setlength{\tabcolsep}{0.2em} 
	\begin{tabular}{cc|cc}
    \multicolumn{2}{c}{\textbf{Fixed Mask}}
    & \multicolumn{2}{c}{\textbf{Variable Mask}}
    \\ 
    \cmidrule(r){1-2} \cmidrule(r){3-4} 
    \includegraphics[width=\figsizeunet\linewidth,valign=m]{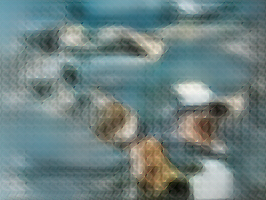}
    &
    \includegraphics[width=\figsizeunet\linewidth,valign=m]{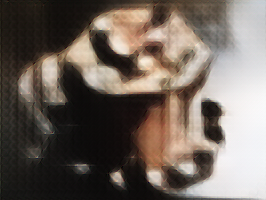}& 
    \includegraphics[width=\figsizeunet\linewidth,valign=m]{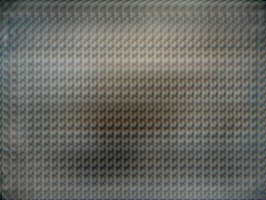}
    &
\includegraphics[width=\figsizeunet\linewidth,valign=m]{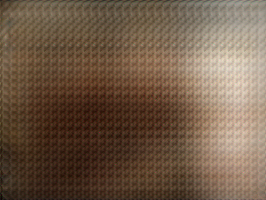}
    \\
	\end{tabular}
	\caption{Example reconstructions when training a decoder on a leaked dataset: (left) shows the vulnerability of a fixed mask pattern and (right) the defense of a variable mask pattern.}
    \label{fig:exp3_privacy}
\end{figure}

Adversaries could also attempt to train a neural network decoder using a dataset of lensless-original image pairs. This could occur as part of a CPA, where the adversary controls the input plaintexts, or as a KPA, 
where image pairs are obtained via a compromised camera or data leak. 
A programmable system provides an effective defense against such attacks, as training a network to invert multiple mask patterns is considerably more difficult than training for a single mask.
To evaluate this,
we train a DRUNet with 8M parameters on \textit{MirFlickr-S} and \textit{MirFlickr-M},
namely the adversary trains a decoder where the PSF (secret key) is not needed as they would not have access to this for a new image.
Example reconstructions are shown in \cref{fig:exp3_privacy}, and the corresponding average image quality metrics are reported in \cref{tab:exp3_privacy}. 
As expected, systems with a fixed mask (\textit{MirFlickr-S}) are vulnerable to CPA and KPA-based attacks, enabling adversaries to reconstruct discernible images without knowledge of the PSF (left-most images of \cref{fig:exp3_privacy}). 
In contrast, systems using variable masks (\textit{MirFlickr-M}) effectively thwart such attacks, as reconstructions are indecipherable, with more than \SI{4}{\decibel} reduction in PSNR.
On the other hand, legitimate users should be able to recover meaning outputs with the known PSF. 
While ADMM reconstructions are suboptimal, 
employing \textit{Learned} is significantly better,
as seen in \cref{tab:exp_multimask} and \cref{fig:exp1_admm}.
In summary, variable-mask imaging systems offer robust defenses against CPAs and KPAs by preventing adversaries from recovering meaningful images or training effective decoders. 

\begin{table}[!t]
\renewcommand{\arraystretch}{1.3}
\centering
\caption{Average image quality metrics of training a DRUNet decoder when the mask is fixed (\textit{MirFlickr-S}) and when the mask pattern varies (\textit{MirFlickr-M}).}
\label{tab:exp3_privacy}
\begin{tabular}{|c|c|c|c|}
\hline
& \textbf{PSNR $\uparrow$} & \textbf{SSIM $\uparrow$} & \textbf{LPIPS $\downarrow$} \\
\hline
MirFlickr-S & 17.9 & 0.459 & 0.511 \\
\hline
MirFlickr-M & 13.7  & 0.274 & 0.587 \\
\hline
\end{tabular}
\end{table}

\begin{table}[!t]
\renewcommand{\arraystretch}{1.3}
\centering
\caption{Average image quality metrics when the mask pattern varies, \ie test set of \textit{MirFlickr-M}.}
\label{tab:exp_multimask}
\begin{tabular}{|c|c|c|c|}
\hline
& \textbf{PSNR $\uparrow$} & \textbf{SSIM $\uparrow$} & \textbf{LPIPS $\downarrow$} \\
\hline
ADMM &  10.6 & 0.301 & 0.760 \\
\hline
Learned & \textbf{18.5}  & \textbf{0.509} & \textbf{0.477} \\
\hline
\end{tabular}
\end{table}

\subsection{Encryption Strength Against Brute-Force Attacks}
\label{sec:encryption_exp}

\begin{figure*}[t!]
\centering
\begin{subfigure}{0.24\linewidth}
    \centering
    \includegraphics[width=0.99\linewidth]{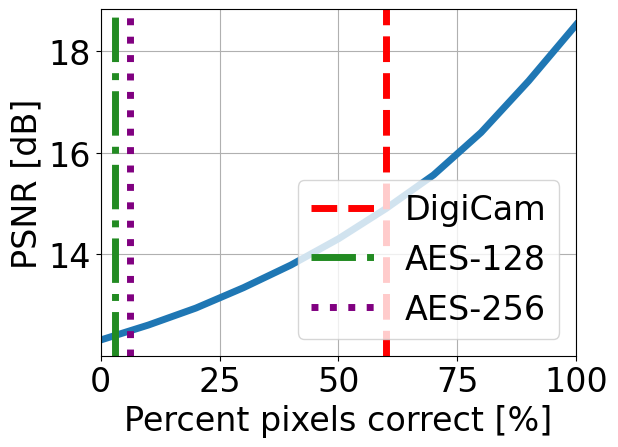} 
    \caption{PSNR.}
    \label{fig:psnr}
\end{subfigure}
\begin{subfigure}{0.24\linewidth}
    \centering
    \includegraphics[width=0.99\linewidth]{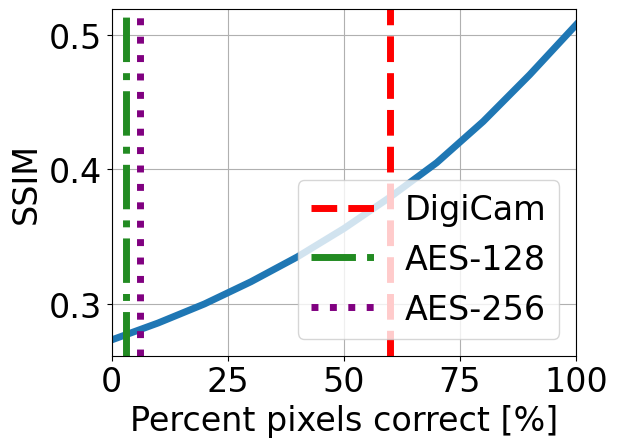} 
    \caption{SSIM.}
    \label{fig:ssim}
\end{subfigure}
\begin{subfigure}{0.24\linewidth}
    \centering
    \includegraphics[width=0.99\linewidth]{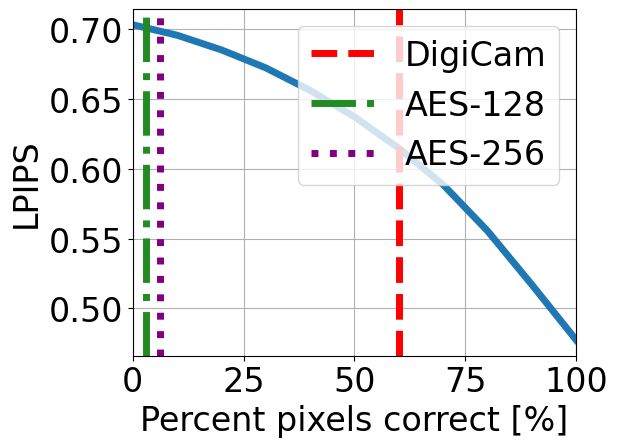}
    \caption{LPIPS.}
    \label{fig:lpips}
\end{subfigure}
\begin{subfigure}{0.24\linewidth}
    \centering
    \includegraphics[width=0.99\linewidth]{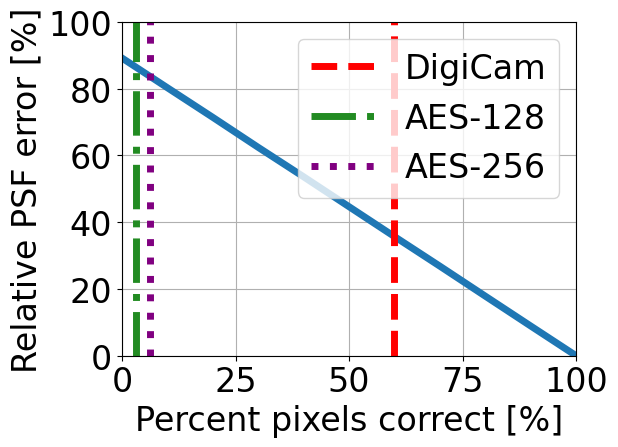}
    \caption{Relative PSF error.}
    \label{fig:psf_err}
\end{subfigure}
\caption{\cref{fig:psnr,fig:ssim,fig:lpips} plot average image quality metrics (PSNR $\uparrow$, SSIM $\uparrow$, LPIPS $\downarrow$) for a modular learned reconstruction~\cite{Bezzam2024} as the percentage of correct mask pattern varies. \cref{fig:psf_err} shows the relative error (percentage) in the PSF. The vertical lines indicate the search space for different encryption approaches.
}
\label{fig:exp1_improvement_viz}
\end{figure*}

\newcommand{\figsizepsferr}{0.12}
\newcommand{\newlinepsfeff}{18pt}
\begin{figure*}[t!]
\centering
	\renewcommand{\arraystretch}{1} 
	\setlength{\tabcolsep}{0.06em} 
	\begin{tabular}[b]{ccccccc}

    & \SI{0}{\percent} &\SI{20}{\percent}&\SI{40}{\percent}&\SI{60}{\percent}&\SI{80}{\percent}&\SI{100}{\percent}\\

\makecell{Example PSF with\\\% correct pixels $\rightarrow$\\ \rule{70pt}{0.5pt} \\ Ground-truth $\downarrow$} &
\includegraphics[width=\figsizepsferr\linewidth,valign=m]{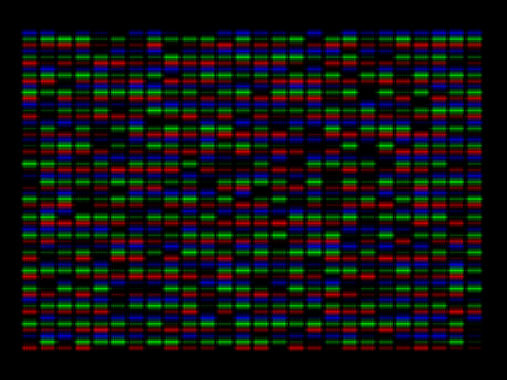}
&\includegraphics[width=\figsizepsferr\linewidth,valign=m]{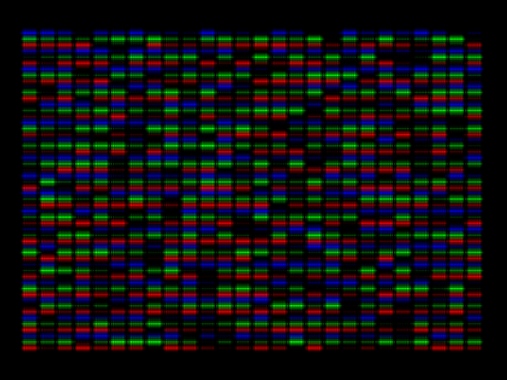}
&\includegraphics[width=\figsizepsferr\linewidth,valign=m]{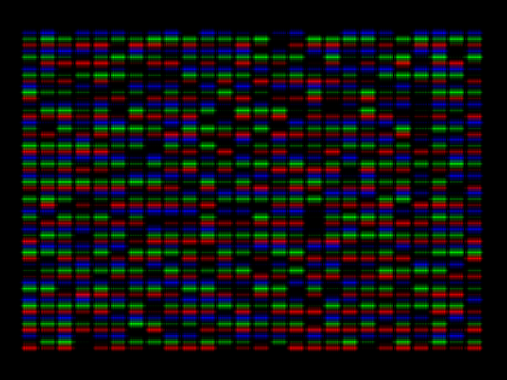}
&\includegraphics[width=\figsizepsferr\linewidth,valign=m]{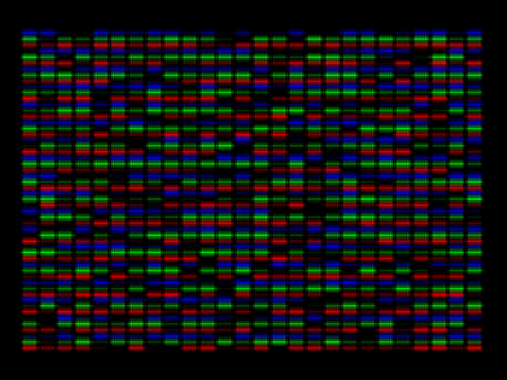}
&\includegraphics[width=\figsizepsferr\linewidth,valign=m]{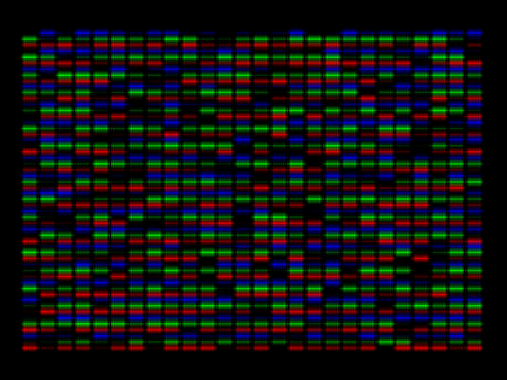}
&\includegraphics[width=\figsizepsferr\linewidth,valign=m]{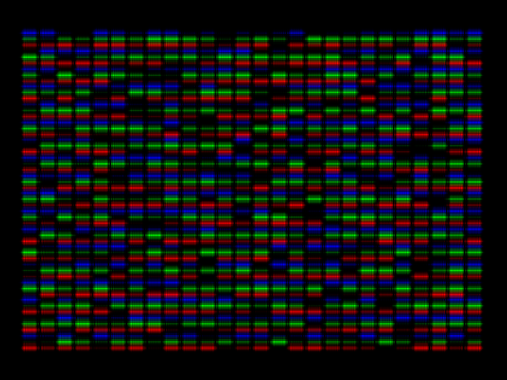}
    \\[\newlinepsfeff]

\hline
\includegraphics[width=\figsizepsferr\linewidth,valign=m]{figs/Unet4M+U5+Unet4M_wave_psfNN/1/original_idx1.png}
&
\includegraphics[width=\figsizepsferr\linewidth,valign=m]{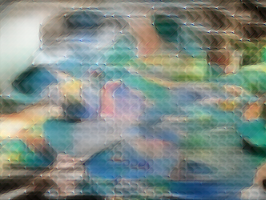}
&\includegraphics[width=\figsizepsferr\linewidth,valign=m]{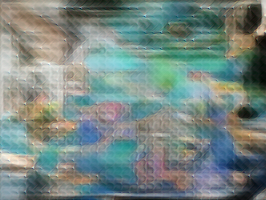}
&\includegraphics[width=\figsizepsferr\linewidth,valign=m]{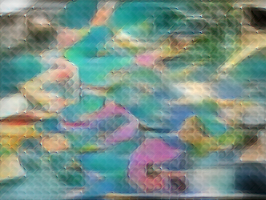}
&\includegraphics[width=\figsizepsferr\linewidth,valign=m]{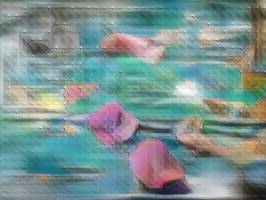}
&\includegraphics[width=\figsizepsferr\linewidth,valign=m]{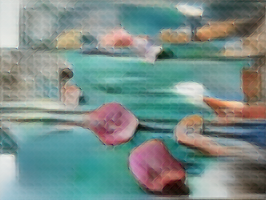}
&\includegraphics[width=\figsizepsferr\linewidth,valign=m]{figs/Unet4M+U5+Unet4M_wave_psfNN/1/Unet4M+U5+Unet4M_wave_psfNN_percentwrong0.png}
    \\[\newlinepsfeff]
\includegraphics[width=\figsizepsferr\linewidth,valign=m]{figs/Unet4M+U5+Unet4M_wave_psfNN/2/original_idx2.png}
&
\includegraphics[width=\figsizepsferr\linewidth,valign=m]{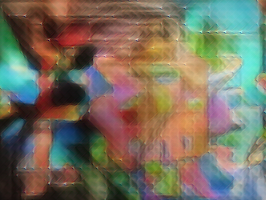}
&\includegraphics[width=\figsizepsferr\linewidth,valign=m]{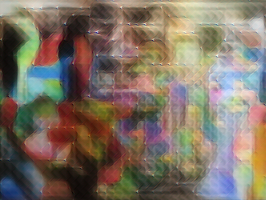}
&\includegraphics[width=\figsizepsferr\linewidth,valign=m]{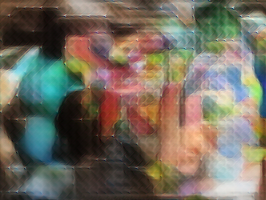}
&\includegraphics[width=\figsizepsferr\linewidth,valign=m]{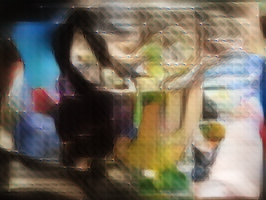}
&\includegraphics[width=\figsizepsferr\linewidth,valign=m]{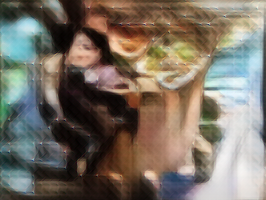}
&\includegraphics[width=\figsizepsferr\linewidth,valign=m]{figs/Unet4M+U5+Unet4M_wave_psfNN/2/Unet4M+U5+Unet4M_wave_psfNN_percentwrong0.png}
    \\[\newlinepsfeff]
\includegraphics[width=\figsizepsferr\linewidth,valign=m]{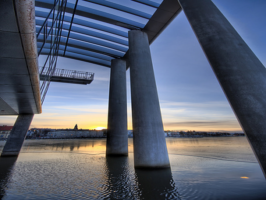}
&
\includegraphics[width=\figsizepsferr\linewidth,valign=m]{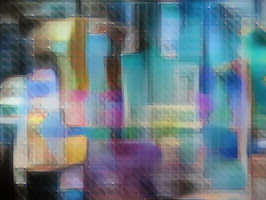}
&\includegraphics[width=\figsizepsferr\linewidth,valign=m]{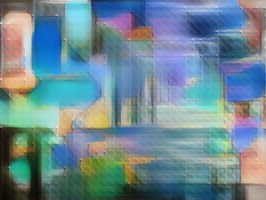}
&\includegraphics[width=\figsizepsferr\linewidth,valign=m]{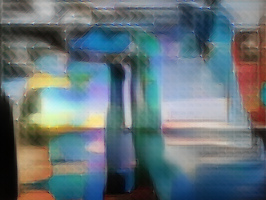}
&\includegraphics[width=\figsizepsferr\linewidth,valign=m]{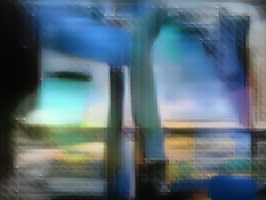}
&\includegraphics[width=\figsizepsferr\linewidth,valign=m]{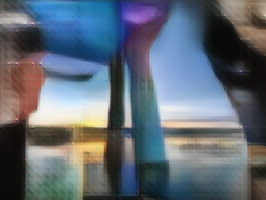}
&\includegraphics[width=\figsizepsferr\linewidth,valign=m]{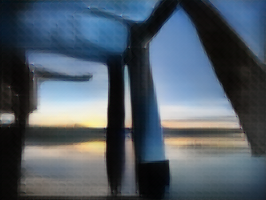}
    \\
\end{tabular}
	\caption{Example reconstructions with a modular learned reconstruction~\cite{Bezzam2024} as the percentage of correct pixels in the mask pattern increases. Top row shows example PSFs according to the percentage of correct pixels.}
    \label{fig:exp2_encrypt}
\end{figure*}

In the previous experiment, we demonstrated that using an incorrect PSF produces distorted reconstructions (\cref{fig:exp1_admm}). 
However, an adversary might attempt a brute-force attack, 
testing multiple mask combinations until achieving a discernible reconstruction. 
This scenario represents a COA, where the adversary only has access to the encrypted data.
The \textit{DigiCam} system used in this work has $N = 1404$ controllable pixels of bit-depth of $b=8$.
Using \cref{eq:correct_bound}, 
the lower bound on the proportion of correct mask pixels required to match or exceed the security of standard encryption key lengths $K\in[128,256]$ is $W \in [0.03,0.06]$.
Therefore,
to achieve comparable security to AES-128 and AES-256, 
more than \SI{3}{\percent} and \SI{6}{\percent} of the mask pixels, respectively, must be correct.
This experiment investigates the empirical threshold of correct mask pixels needed for discernible reconstructions.
To determine this, we vary the proportion of altered pixels in the test set masks from \textit{MirFlickr-M}. 
For each test example,
we randomly perturb a fixed percentage of mask pixels, simulate the corresponding PSF, and perform image recovery \textit{Learned}.
Example reconstructions in \cref{fig:exp2_encrypt} demonstrate that significantly more than \SI{6}{\percent} of correct pixels are required for a meaningful image reconstruction.
Discernible images emerge when approximately $W=\SI{60}{\percent}$ of the mask pixels are correct,
corresponding to an effective binary key length of $K=2'527$ bits based on \cref{eq:correct_bound}.
This far surpasses the security of AES-256 and demonstrates the robustness of \textit{DigiCam} against brute-force COAs.
As shown in \cref{fig:psnr,fig:ssim,fig:lpips}, the average image quality declines as the proportion of correct mask pixels decreases (to the left).
\cref{fig:psf_err} approximates the deviation $\bm{\Delta}$ in \cref{eq:psf_err} as the relative PSF error,
offering insight into the required deviation to thwart decryption of encrypted data.

\subsection{Authentication Capabilities}
\label{sec:auth_exp}

\begin{table}[!t]
\renewcommand{\arraystretch}{1.3}
\centering
\caption{Average authentication accuracy when taking the minimum score across 15 masks from \textit{DigiCam-M} dataset.}
\label{tab:exp3_auth}
\begin{tabular}{|c|c|c|c|}
\hline 
& \makecell{\textbf{Data Fidelity}\\\textbf{(no lensed data)}} & \textbf{MSE}  & \textbf{LPIPS} \\
\hline
ADMM & \SI{79.3}{\percent} & \SI{51.9}{\percent} & \SI{96.5}{\percent} \\
\hline
Learned & \SI{81.4}{\percent} & \SI{96.6}{\percent} & \SI{99.5}{\percent} \\
\hline
\end{tabular}
\end{table}

As discussed in \cref{sec:method_auth}, a unique mask pattern can be generated for a specific user or camera and timestamp, and a variety of scores can validate a measurement against its associated mask pattern.
Using data fidelity, MSE, and LPIPS as authentication scores with \cref{eq:auth,eq:auth_ref}, 
\cref{tab:exp3_auth} presents average authentication accuracy for the test set of \textit{MirFlickr-M}, 
where each measurement is tested against the 15 mask patterns in the test set.
The \textit{Learned} approach outperforms ADMM, achieving near-perfect accuracy with the LPIPS score.
While ADMM has the advantage of being training-free, 
using MSE as score suffers
due to discoloration effects (see \cref{fig:exp1_admm}), 
as pixelwise metrics are highly sensitive to color shifts. 
Conversely, the LPIPS score is more robust, operating perceptually with patchwise comparisons in a neural network feature space.

\begin{figure}[t!]
\centering
\begin{subfigure}{0.49\linewidth}
    \centering
    \includegraphics[width=0.99\linewidth]{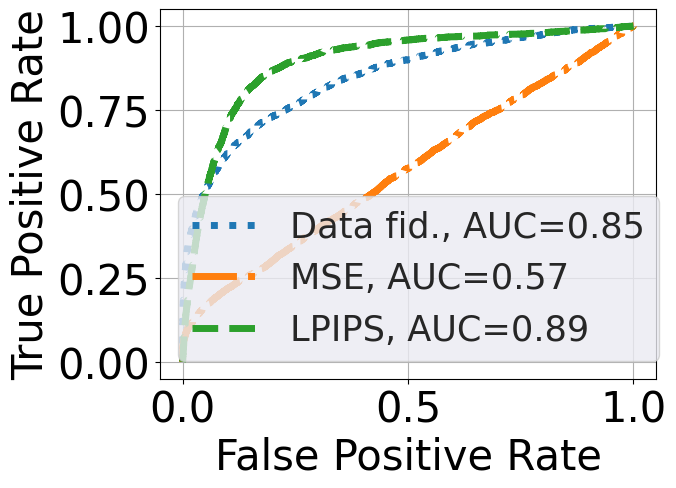} 
    \caption{ADMM.}
    \label{fig:roc_curve_admm100}
\end{subfigure}
\begin{subfigure}{0.49\linewidth}
    \centering
    \includegraphics[width=0.99\linewidth]{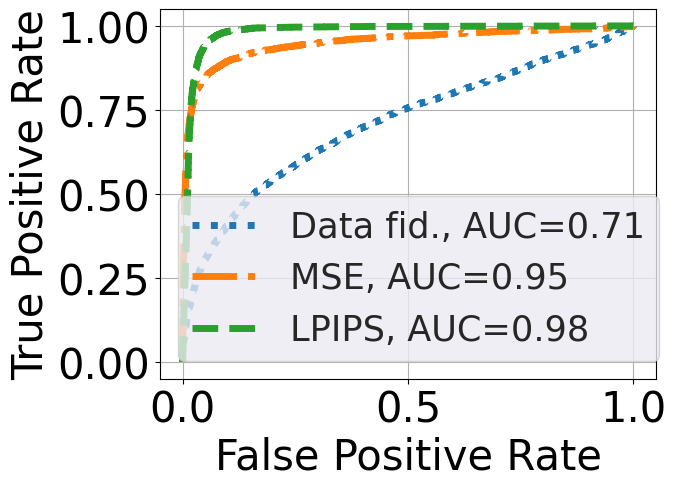} 
    \caption{Learned.}
    \label{fig:roc_curve_learned}
\end{subfigure}
\caption{\textit{Receiver Operating Characteristic} curves for authentication with \cref{eq:auth,eq:auth_ref}.}
\label{fig:exp3_roc}
\end{figure}

For measurements with a new mask pattern, a binary threshold on the authentication score is required.
\cref{fig:exp3_roc} illustrates Receiver Operating Characteristic (ROC) curves 
where a threshold is varied on the multiple authentication scores, 
and the area under the curve (AUC) quantifies the quality of the detector.
An AUC above 0.9 is considered excellent. 
The results indicate that ADMM performs better for authentication when only the lensless image is available, due to its higher AUC for data fidelity scores. 
However, when validating lensed images with
lensless measurements as certificates (\eg with MSE or LPIPS), the learned reconstruction approach is more advantageous (see \cref{fig:roc_curve_learned}).

In summary, using a programmable mask, as in \textit{DigiCam}, offers a robust framework for image verification,
enabling the validation of when and by whom an image was captured through a candidate mask pattern.
Performance is enhanced when combined with a lensed image in a multi-sensor setup.
In such cases, the lensless measurement (and its mask pattern) serve as an analog certificate, 
offering a reliable method to establish the authenticity of an image or video.

\section{Conclusion}
\label{sec:conclusion}

In this work, we enhanced the security capabilities of a lensless cameras by using a programmable mask that dynamically varies its pattern to thwart adversarial attacks. Additionally, we demonstrated a novel application in image authentication, leveraging the unique fingerprints left by each mask.
For future work, further investigation into the structured nature of lensless encryption could uncover more efficient search strategies for adversaries.
For authentication, optimizing the mask generation process by identifying maximally separable patterns could improve accuracy.
Moreover, applications in video and edge devices could be considered.

\subsubsection*{Acknowledgement} This work was supported by the Swiss National Science
Foundation under Grant CRSII5\textunderscore213521 ``DigiLight---Programmable Third-Harmonic Generation (THG) Microscopy Applied to Advanced Manufacturing''.

\bibliographystyle{IEEEbib}
\bibliography{main}

\begin{thebibliography}{10}

\bibitem{Javidi_2016}
Bahram Javidi et~al.,
\newblock ``Roadmap on optical security,''
\newblock {\em Journal of Optics}, vol. 18, no. 8, pp. 083001, jul 2016.

\bibitem{boominathan2022recent}
Vivek Boominathan, Jacob~T Robinson, Laura Waller, and Ashok Veeraraghavan,
\newblock ``Recent advances in lensless imaging,''
\newblock {\em Optica}, vol. 9, no. 1, pp. 1--16, 2022.

\bibitem{Tan2020}
Jasper Tan et~al.,
\newblock ``{CANOPIC: Pre-digital privacy-enhancing encodings for computer
  vision},''
\newblock in {\em IEEE Int. Conf. Multimedia and Expo}, Jul. 2020, pp. 1--6.

\bibitem{10666814}
Salman~S. Khan, Xiang Yu, Kaushik Mitra, Manmohan Chandraker, and Francesco
  Pittaluga,
\newblock ``{OpEnCam: Lensless Optical Encryption Camera},''
\newblock {\em IEEE Trans. on Comput. Imaging}, vol. 10, pp. 1306--1316, 2024.

\bibitem{phlatcam}
Vivek Boominathan, Jesse~K. Adams, Jacob~T. Robinson, and Ashok Veeraraghavan,
\newblock ``{PhlatCam: Designed Phase-Mask Based Thin Lensless Camera},''
\newblock {\em IEEE Trans. Pattern Anal. Mach. Intell.}, vol. 42, no. 7, pp.
  1618--1629, 2020.

\bibitem{Lee:23}
Kyung~Chul Lee et~al.,
\newblock ``Design and single-shot fabrication of lensless cameras with
  arbitrary point spread functions,''
\newblock {\em Optica}, vol. 10, no. 1, pp. 72--80, Jan 2023.

\bibitem{1634362}
J.~Lukas, J.~Fridrich, and M.~Goljan,
\newblock ``Digital camera identification from sensor pattern noise,''
\newblock {\em IEEE Trans. on Information Forensics and Security}, vol. 1, no.
  2, pp. 205--214, 2006.

\bibitem{10.1117/12.649775}
Kai~San Choi, Edmund~Y. Lam, and Kenneth K.~Y. Wong,
\newblock ``{Source camera identification using footprints from lens
  aberration},''
\newblock in {\em Digital Photography II}. 2006, SPIE.

\bibitem{6126225}
Zhonghai Deng, Arjan Gijsenij, and Jingyuan Zhang,
\newblock ``Source camera identification using auto-white balance
  approximation,''
\newblock in {\em Int. Conf. Comput. Vis.}, 2011, pp. 57--64.

\bibitem{Bezzam2024}
Eric Bezzam, Yohann Perron, and Martin Vetterli,
\newblock ``Towards robust and generalizable lensless imaging with modular
  learned reconstruction,''
\newblock {\em IEEE Trans. on Comput. Imaging}, vol. 11, pp. 213--227, 2025.

\bibitem{Ryoo2017}
Michael Ryoo, Brandon Rothrock, Charles Fleming, and Hyun~Jong Yang,
\newblock ``{Privacy-preserving human activity recognition from extreme low
  resolution},''
\newblock {\em Proc. of the AAAI Conf. Artif. Intell.}, vol. 31, no. 1, Feb.
  2017.

\bibitem{Hinojosa2022}
Carlos Hinojosa et~al.,
\newblock ``{PrivHAR: Recognizing human actions from privacy-preserving
  lens},''
\newblock in {\em Eur. Conf. Comput. Vis.}, 2022, pp. 314--332.

\bibitem{Refregier:95}
Philippe Refregier and Bahram Javidi,
\newblock ``Optical image encryption based on input plane and fourier plane
  random encoding,''
\newblock {\em Opt. Lett.}, vol. 20, no. 7, pp. 767--769, Apr 1995.

\bibitem{9025454}
Zihao~W. Wang et~al.,
\newblock ``{Privacy-Preserving Action Recognition Using Coded Aperture
  Videos},''
\newblock in {\em IEEE Conf. Comput. Vis. Pattern Recog. Workshops (CVPRW)},
  2019, pp. 1--10.

\bibitem{flatcam}
M.~Salman Asif, Ali Ayremlou, Aswin Sankaranarayanan, Ashok Veeraraghavan, and
  Richard~G. Baraniuk,
\newblock ``{FlatCam: Thin, Lensless Cameras Using Coded Aperture and
  Computation},''
\newblock {\em IEEE Trans. on Comput. Imaging}, vol. 3, no. 3, pp. 384--397,
  2017.

\bibitem{ADMM}
Stephen Boyd, Neal Parikh, Eric Chu, Borja Peleato, and Jonathan Eckstein,
\newblock ``Distributed optimization and statistical learning via the
  alternating direction method of multipliers,''
\newblock {\em Foundations and Trends in Mach. Learning}, vol. 3, no. 1, pp.
  1--122, 2011.

\bibitem{Antipa:18}
Nick Antipa et~al.,
\newblock ``{DiffuserCam: lensless single-exposure 3D imaging},''
\newblock {\em Optica}, vol. 5, no. 1, pp. 1--9, Jan 2018.

\bibitem{Monakhova:19}
Kristina Monakhova et~al.,
\newblock ``Learned reconstructions for practical mask-based lensless
  imaging,''
\newblock {\em Opt. Express}, vol. 27, no. 20, pp. 28075--28090, Sep 2019.

\bibitem{9239993}
Salman~S. Khan, Varun Sundar, Vivek Boominathan, Ashok Veeraraghavan, and
  Kaushik Mitra,
\newblock ``{FlatNet: Towards Photorealistic Scene Reconstruction from Lensless
  Measurements},''
\newblock {\em IEEE Trans. Pattern Anal. Mach. Intell.}, Oct 2020.

\bibitem{Li:23}
Ying Li, Zhengdai Li, Kaiyu Chen, Youming Guo, and Changhui Rao,
\newblock ``{MWDN}s: reconstruction in multi-scale feature spaces for lensless
  imaging,''
\newblock {\em Opt. Express}, vol. 31, no. 23, pp. 39088--39101, Nov 2023.

\bibitem{Kishk:02}
Sherif Kishk and Bahram Javidi,
\newblock ``Information hiding technique with double phase encoding,''
\newblock {\em Appl. Opt.}, vol. 41, no. 26, pp. 5462--5470, Sep 2002.

\bibitem{9546648}
Tianjiao Zeng and Edmund~Y. Lam,
\newblock ``Robust reconstruction with deep learning to handle model mismatch
  in lensless imaging,''
\newblock {\em IEEE Trans. on Comput. Imaging}, vol. 7, pp. 1080--1092, 2021.

\bibitem{zhang2018perceptual}
Richard Zhang, Phillip Isola, Alexei~A Efros, Eli Shechtman, and Oliver Wang,
\newblock ``The unreasonable effectiveness of deep features as a perceptual
  metric,''
\newblock in {\em IEEE Conf. Comput. Vis. Pattern Recog.}, 2018.

\bibitem{huiskes2008mir}
Mark~J Huiskes and Michael~S Lew,
\newblock ``The {MIR F}lickr retrieval evaluation,''
\newblock in {\em Proceedings ACM Int. Conf. on Multimedia Information
  Retrieval}, 2008, pp. 39--43.

\bibitem{zhang2021plug}
Kai Zhang et~al.,
\newblock ``Plug-and-play image restoration with deep denoiser prior,''
\newblock {\em IEEE Trans. Pattern Anal. Mach. Intell.}, vol. 44, no. 10, pp.
  6360--6376, 2021.

\bibitem{adafruitlcd}
``{1.8" Color TFT LCD display with MicroSD Card Breakout - ST7735R},''
  \url{https://www.adafruit.com/product/358} (Aug. 2024).

\bibitem{Matsushima2009}
Kyoji Matsushima and Tomoyoshi Shimobaba,
\newblock ``{Band-limited angular spectrum method for numerical simulation of
  free-space propagation in far and near fields},''
\newblock {\em Optics Express}, vol. 17, no. 22, pp. 19662, Oct. 2009.

\end{thebibliography}

\crefalias{section}{appendix}

\appendices

\renewcommand\thefigure{\thesection.\arabic{figure}}  
\setcounter{figure}{0}   
\renewcommand{\theequation}{\thesection.\arabic{equation}}
\setcounter{equation}{0}   
\renewcommand{\thetable}{\thesection.\arabic{table}}
\setcounter{table}{0}   

\section{Deviation due to Decoding with Wrong PSF}
\label{app:deviation}

We can express the image capture process as a matrix-vector product with a system matrix $\bm{H}$ with additive noise:
\begin{align}
	\bm{y} = \bm{Hx} + \bm{n}.
\end{align}
If the measurement is decoded with a different system matrix $\hat{\bm{H}}=(\bm{H}-\bm{\Delta})$,
applying direct inversion with this system matrix yields:
\begin{align}
	\bm{\hat{x}} &= \hat{\bm{H}}^{-1}\bm{y} \\
	&= (\bm{H}-\bm{\Delta})^{-1} (\bm{H}\bm{x} + \bm{n})  \\
	&= (\bm{I}-\bm{H}^{-1}\bm{\Delta})^{-1} (\bm{x} + \bm{H}^{-1}\bm{n})  \\
	&= \bm{x} + \bm{H}^{-1}\bm{n} + \underbrace{( \bm{x} + \bm{H}^{-1}\bm{n})\sum_{k=1}^{\infty}(\bm{H}^{-1}\bm{\Delta})^k}_{\text{error when using wrong system}} \label{eq:psf_err_app},
\end{align}
where the last line uses the Taylor expansion $(\bm{I}-\bm{A})^{-1} = \bm{I} + \sum_{k=1}^{\infty} \bm{A}^k$ with $\bm{A} = \bm{H}^{-1}\bm{\Delta}_H$.
With \cref{eq:psf_err_app}, we see that
the larger the deviation $\bm{\Delta}$ from the the true system,
the larger the resulting error.

\section{Point Spread Function Simulation}

As point spread function (PSF) simulation has been extensily described in~\cite{Bezzam2024}, we provide a brief overview and describe necessary details here. 
In this work, 
we use the \textit{DigiCam} prototype,
which makes use of a low-cost LCD~\cite{adafruitlcd} (around $20$ USD).
As knowledge of the mask pattern (secret key) set to the LCD,
the PSF can be simulated as such:
\begin{align}
	\label{eq:intensity_psf_simple}
	&p(\bm{r}, c; d_1, d_2, \bm{w}) = \nonumber \\
	&|\mathcal{F}^{-1}\Big(\mathcal{F} \Big( m(\bm{r}, \lambda_c; \bm{w}) e^{j \frac{2\pi}{\lambda_c} \sqrt{\|\bm{r}\|_2^2 +  d_1^2}}
	\Big) \times h(\bm{u}, \lambda_c; z=d_2) \Big)|^2,
\end{align}
where $\bm{r} \in \mathbb{R}^2$ are positions along a 2D plane, 
$ c\in\{R,G,B\} $ are the color channels whose corresponding wavelengths are $ (\lambda_R, \lambda_G, \lambda_B) = (\SI{640}{\nano\meter}, \SI{550}{\nano\meter}, \SI{460}{\nano\meter}) $,
$d_1$ is the distance from the scene to the LCD,
$d_2$ is the distance from the LCD to the sensor,
$\bm{w}\in [0, 1]^N$ are the pixel weights to construct the mask pattern $m(\bm{r}; \lambda, \bm{w})$,
$\mathcal{F}$ and $\mathcal{F}^{-1}$ refer to the 2D Fourier transform and its inverse respectively,
$ h(\bm{u}; z, \lambda)$ is the free-space propagation frequency response~\cite{Matsushima2009},
and $\bm{u} \in \mathbb{R}^2$ are the spatial frequencies of $\bm{r}$.
The mask pattern $m(\bm{r}; \lambda, \bm{w})$ for a programmable mask can be modeled as a superposition of apertures for each adjustable RGB sub-pixel in $ \bm{r} \in \mathbb{R}^2 $~\cite{Bezzam2024}.

\begin{figure}[t!]
	\centering
	\includegraphics[width=0.9\linewidth]{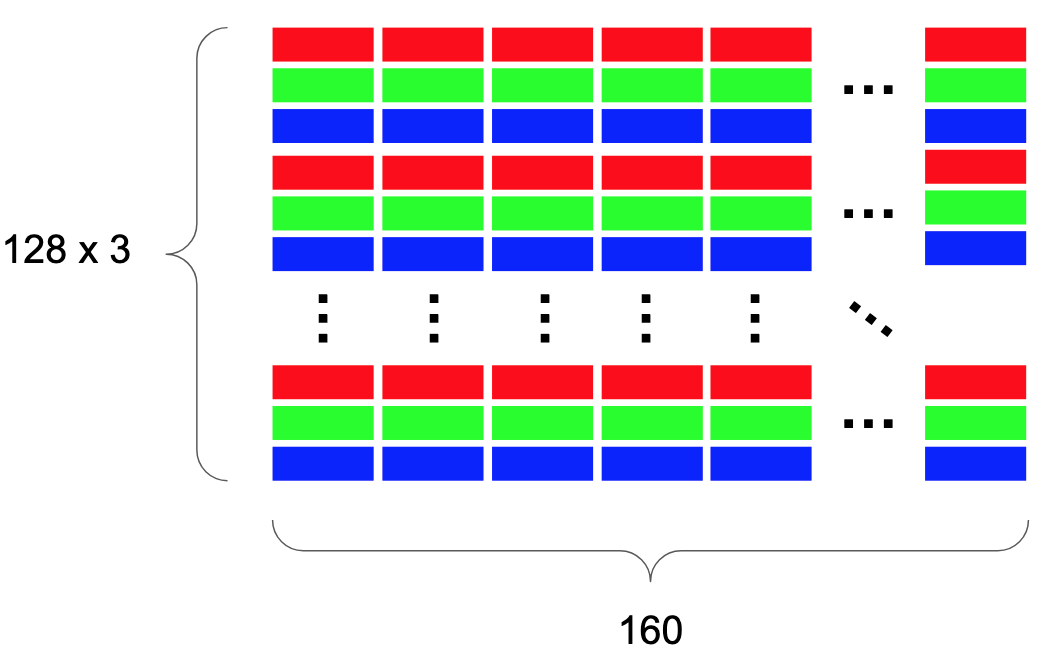}
	\caption{Pixel layout of the ST7735R component~\cite{adafruitlcd}: red, green, blue color filter arrangement.}
	\label{fig:pixel_layout}
\end{figure}

As the chosen LCD component in our lensless camera is color display,
it has an interleaved pattern of red, blue, and green sub-pixels,
as shown in \cref{fig:pixel_layout}.
We only use a subset of the pixels that overlap with the sensor,
\ie $(3\times18\times26) = 1404$ sub-pixels.

\section{DigiCam System}

\begin{figure}[t!]
	\centering
	\includegraphics[width=0.9\linewidth]{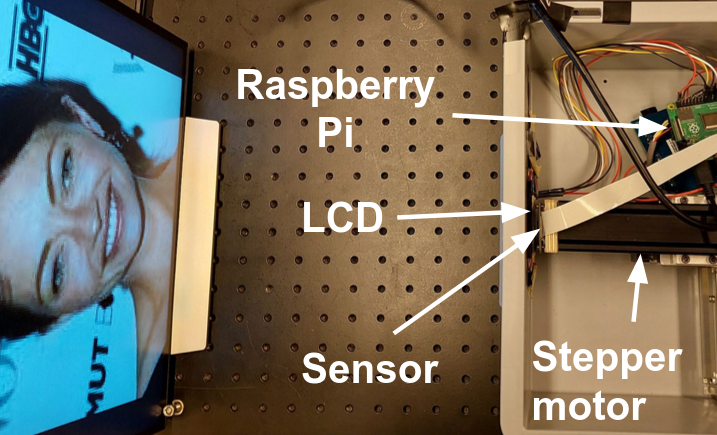}
	\caption{DigiCam prototype and measurement setup.}
	\label{fig:prototype_labeled}
\end{figure}

The \textit{DigiCam} prototype and measurement setup can be seen in \cref{fig:prototype_labeled}, where a digital monitor is placed \SI{30}{\centi\meter} from the lensless camera to display content during data acquisition.
The LCD is placed \SI{2}{\milli\meter} from the sensor.
The prototype includes an optional stepper motor for programmatically setting the distance between the LCD and the sensor.

\section{Encryption Strength against Brute-Force Attacks when using ADMM}

\begin{table*}[!t]
	\renewcommand{\arraystretch}{1.3}
	\centering
	\caption{Average PSNR (in decibels) when reconstructing images from the \textit{MirFlickr-M} test set as the percentage of correct pixels in the mask pattern varies. Higher is better.}
	\label{tab:compare_psnr}
	\begin{tabular}{|c|c|c|c|c|c|c|c|c|c|c|c|c|}
		\hline 
		&\multicolumn{11}{|c|}{\textbf{\% of Correct Pixels in Mask Pattern}} \\
		\cline{2-12} 
		\textbf{Decoder} & \SI{0}{\percent} & \SI{10}{\percent}  & \SI{20}{\percent} & \SI{30}{\percent} & \SI{40}{\percent}  & \SI{50}{\percent}& \SI{60}{\percent} & \SI{70}{\percent}  & \SI{80}{\percent}& \SI{90}{\percent}  & \SI{100}{\percent}\\
		\hline
		ADMM & 9.81 & 9.96 & 10.1 & 10.2 & 10.4 & 10.4 &
		10.5 & 10.5 & 10.6 &  10.6 & 10.6 \\
		\hline
		Learned & \textbf{12.3} & \textbf{12.6} & \textbf{12.9} & \textbf{13.3} &  \textbf{13.8} & \textbf{14.3} &
		\textbf{14.9} & \textbf{15.6} & \textbf{16.4} & \textbf{17.4} & \textbf{18.5} \\
		\hline
	\end{tabular}
\end{table*}

\begin{table*}[!t]
	\renewcommand{\arraystretch}{1.3}
	\centering
	\caption{Average SSIM when reconstructing images from the \textit{MirFlickr-M} test set as the percentage of correct pixels in the mask pattern varies. Higher is better.}
	\label{tab:compare_ssim}
	\begin{tabular}{|c|c|c|c|c|c|c|c|c|c|c|c|c|}
		\hline 
		&\multicolumn{11}{|c|}{\textbf{\% of Correct Pixels in Mask Pattern}} \\
		\cline{2-12} 
		\textbf{Decoder} & \SI{0}{\percent} & \SI{10}{\percent}  & \SI{20}{\percent} & \SI{30}{\percent} & \SI{40}{\percent}  & \SI{50}{\percent}& \SI{60}{\percent} & \SI{70}{\percent}  & \SI{80}{\percent}& \SI{90}{\percent}  & \SI{100}{\percent}\\
		\hline
		ADMM & 0.223 &  0.231 & 0.238 &  0.246 & 0.253 & 0.260 &
		0.266 & 0.272 & 0.279 & 0.285 & 0.293 \\
		\hline
		Learned & \textbf{0.273} & \textbf{0.286} & \textbf{0.300} &  \textbf{0.316} & \textbf{0.335} & \textbf{0.356} &
		\textbf{0.379} & \textbf{0.405} & \textbf{0.435} & \textbf{0.470} & \textbf{0.507} \\
		\hline
	\end{tabular}
\end{table*}

\begin{table*}[!t]
	\renewcommand{\arraystretch}{1.3}
	\centering
	\caption{Average LPIPS when reconstructing images from the \textit{MirFlickr-M} test set as the percentage of correct pixels in the mask pattern varies. Lower is better.}
	\label{tab:compare_lpips}
	\begin{tabular}{|c|c|c|c|c|c|c|c|c|c|c|c|c|}
		\hline 
		&\multicolumn{11}{|c|}{\textbf{\% of Correct Pixels in Mask Pattern}} \\
		\cline{2-12} 
		\textbf{Decoder} & \SI{0}{\percent} & \SI{10}{\percent}  & \SI{20}{\percent} & \SI{30}{\percent} & \SI{40}{\percent}  & \SI{50}{\percent}& \SI{60}{\percent} & \SI{70}{\percent}  & \SI{80}{\percent}& \SI{90}{\percent}  & \SI{100}{\percent}\\
		\hline
		ADMM & 0.862 & 0.858 & 0.853 & 0.846 & 0.839 & 0.831 &
		0.821 & 0.810 & 0.797 & 0.781 & 0.760 \\
		\hline
		Learned & \textbf{0.703} & \textbf{0.696} &  \textbf{0.685} & \textbf{0.672} & \textbf{0.656} & \textbf{0.637} &
		\textbf{0.615} & \textbf{0.589} & \textbf{0.556} & \textbf{0.517} & \textbf{0.477} \\
		\hline
	\end{tabular}
\end{table*}

\begin{table*}[!t]
	\renewcommand{\arraystretch}{1.3}
	\centering
	\caption{Average relative point spread function error as the percentage of correct pixels in the mask pattern varies.}
	\label{tab:psf_eff}
	\begin{tabular}{|c|c|c|c|c|c|c|c|c|c|c|c|c|}
		\hline 
		&\multicolumn{11}{|c|}{\textbf{\% of Correct Pixels in Mask Pattern}} \\
		\cline{2-12} 
		\textbf{Decoder} & \SI{0}{\percent} & \SI{10}{\percent}  & \SI{20}{\percent} & \SI{30}{\percent} & \SI{40}{\percent}  & \SI{50}{\percent}& \SI{60}{\percent} & \SI{70}{\percent}  & \SI{80}{\percent}& \SI{90}{\percent}  & \SI{100}{\percent}\\
		\hline
		Relative PSF error & 0.892 & 0.803 & 0.714 & 0.625 & 0.536 & 0.446 &
		0.357 & 0.268 & 0.178 & 0.0892 & 0 \\
		\hline
	\end{tabular}
\end{table*}

In Section~IV-C, we presented image quality metrics (Fig.~5) and example reconstructions (Fig.~6) for the learned reconstruction method as the percent of correct pixels in the mask pattern varies.
In this section, we present the results when ADMM is used as a reconstruction method.
Unlike the learned approach,
this reconstruction does not require training, which would make it more accessible for an adversary,
but the reconstruction quality is much worse (see Fig.~3).
\cref{tab:compare_psnr,tab:compare_ssim,tab:compare_lpips} compare the average image quality metrics between 100 iterations of ADMM and the modular learned reconstruction.
The latter is consistently better.
\cref{tab:psf_eff} shows the average relative PSF error as the percentage of correct pixels varies, which corresponds to Fig.~5d.

\begin{figure*}[t!]
	\centering
	\begin{subfigure}{0.32\linewidth}
		\centering
		\includegraphics[width=0.99\linewidth]{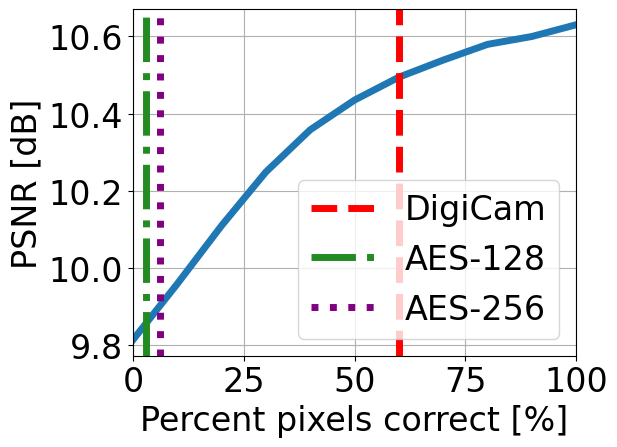} 
		\caption{PSNR.}
		\label{fig:psnr_admm}
	\end{subfigure}
	\begin{subfigure}{0.32\linewidth}
		\centering
		\includegraphics[width=0.99\linewidth]{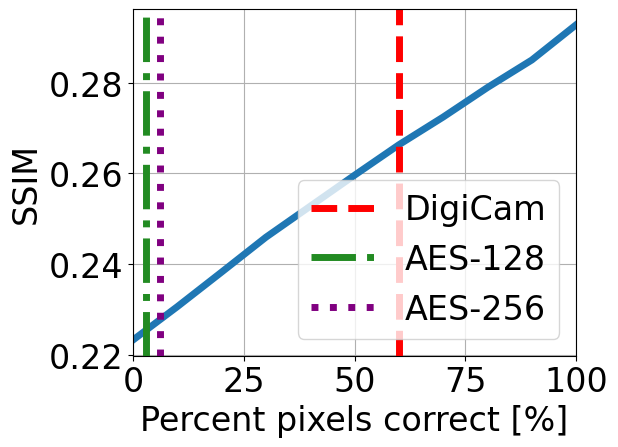} 
		\caption{SSIM.}
		\label{fig:ssim_admm}
	\end{subfigure}
	\begin{subfigure}{0.32\linewidth}
		\centering
		\includegraphics[width=0.99\linewidth]{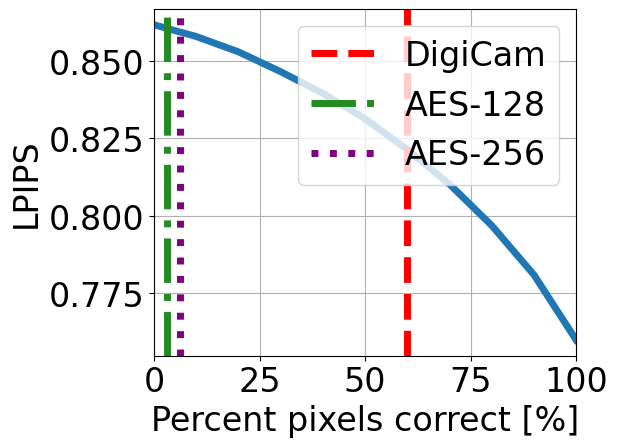}
		\caption{LPIPS.}
		\label{fig:lpips_admm}
	\end{subfigure}
	\caption{(a)--(c) plot average image quality metrics (PSNR $\uparrow$, SSIM $\uparrow$, LPIPS $\downarrow$) when applying 100 iterations of ADMM as the percentage of correct mask pattern varies. The vertical lines indicate the search space for different encryption approaches.
	}
	\label{fig:encryption_admm}
\end{figure*}

For ADMM, \cref{fig:encryption_admm} plots average image quality metrics as percentage of correct pixels increases, and 
\cref{fig:different_admm} shows example reconstructions. 
Similar to the learned approach, 
when around \SI{60}{\percent} of pixels of the mask pattern are correct,
we begin to observe discernible features.

\newcommand{\figsizepsferrapp}{0.14}
\newcommand{\newlinepsfeffapp}{20pt}
\begin{figure*}[t!]
	\centering
	\renewcommand{\arraystretch}{1} 
	\setlength{\tabcolsep}{0.06em} 
	\begin{tabular}[b]{ccccccc}
		
		& \SI{0}{\percent} &\SI{20}{\percent}&\SI{40}{\percent}&\SI{60}{\percent}&\SI{80}{\percent}&\SI{100}{\percent}\\

		\makecell{Example PSF with\\\% correct pixels $\rightarrow$\\ \rule{70pt}{0.5pt} \\ Ground-truth $\downarrow$} &
		\includegraphics[width=\figsizepsferrapp\linewidth,valign=m]{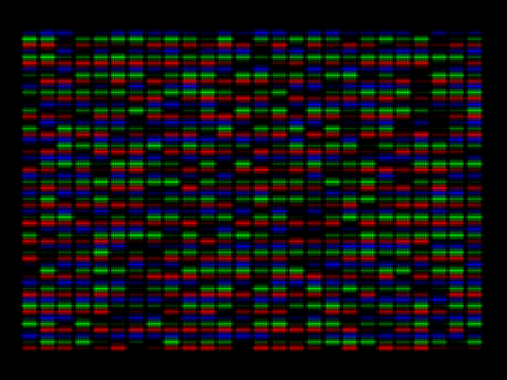}
		&\includegraphics[width=\figsizepsferrapp\linewidth,valign=m]{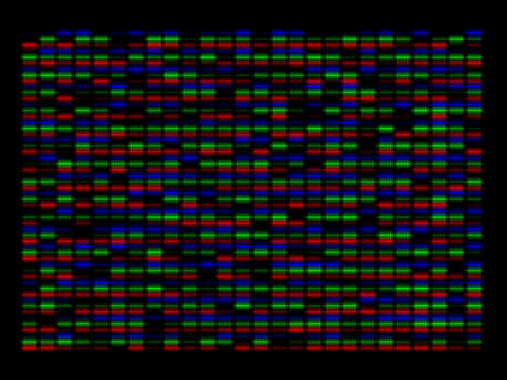}
		&\includegraphics[width=\figsizepsferrapp\linewidth,valign=m]{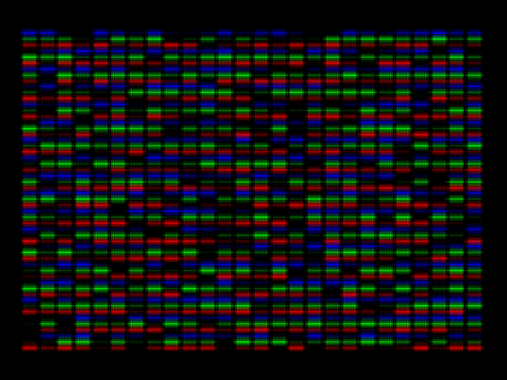}
		&\includegraphics[width=\figsizepsferrapp\linewidth,valign=m]{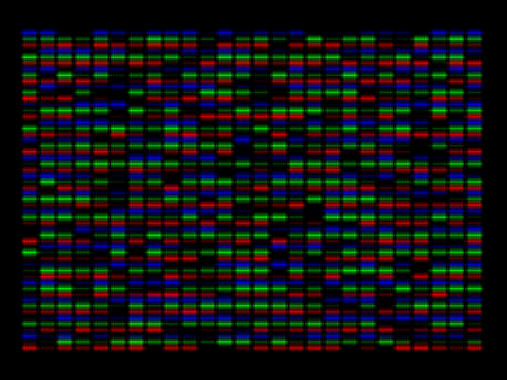}
		&\includegraphics[width=\figsizepsferrapp\linewidth,valign=m]{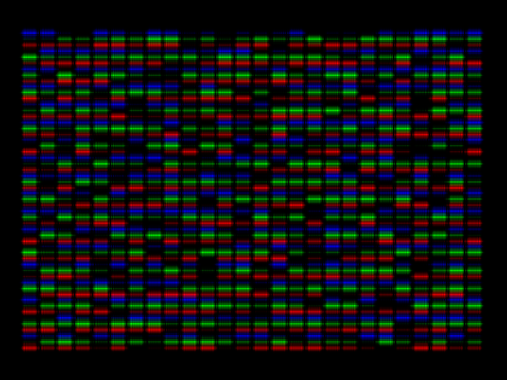}
		&\includegraphics[width=\figsizepsferrapp\linewidth,valign=m]{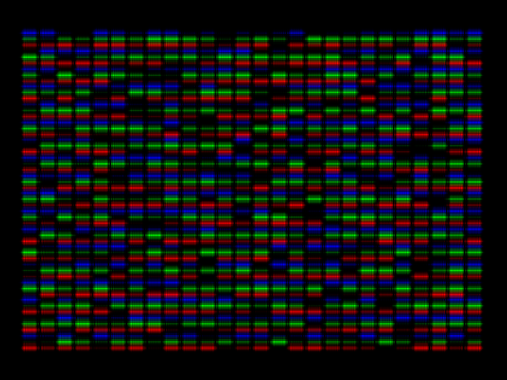}
		\\[\newlinepsfeffapp]
		
		\hline
		\includegraphics[width=\figsizepsferrapp\linewidth,valign=m]{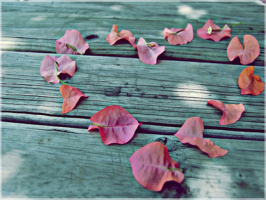}
		&
		\includegraphics[width=\figsizepsferrapp\linewidth,valign=m]{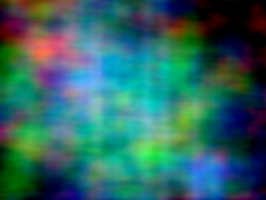}
		&\includegraphics[width=\figsizepsferrapp\linewidth,valign=m]{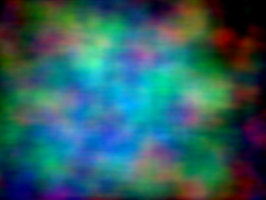}
		&\includegraphics[width=\figsizepsferrapp\linewidth,valign=m]{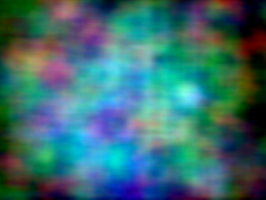}
		&\includegraphics[width=\figsizepsferrapp\linewidth,valign=m]{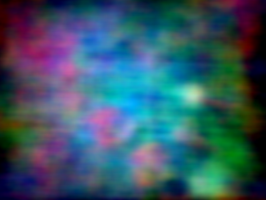}
		&\includegraphics[width=\figsizepsferrapp\linewidth,valign=m]{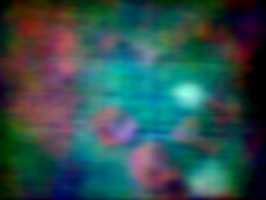}
		&\includegraphics[width=\figsizepsferrapp\linewidth,valign=m]{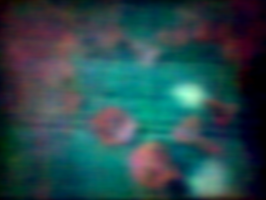}
		\\[\newlinepsfeffapp]
		\includegraphics[width=\figsizepsferrapp\linewidth,valign=m]{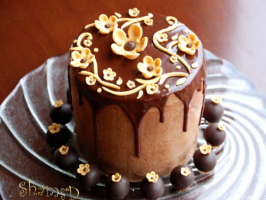}
		&
		\includegraphics[width=\figsizepsferrapp\linewidth,valign=m]{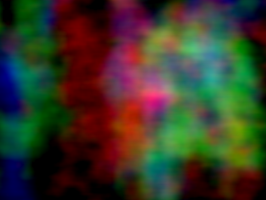}
		&\includegraphics[width=\figsizepsferrapp\linewidth,valign=m]{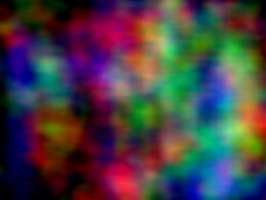}
		&\includegraphics[width=\figsizepsferrapp\linewidth,valign=m]{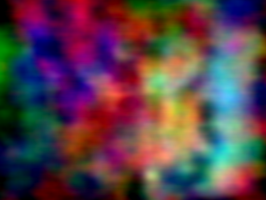}
		&\includegraphics[width=\figsizepsferrapp\linewidth,valign=m]{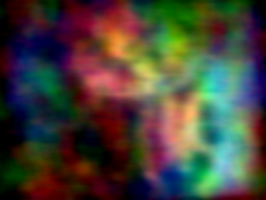}
		&\includegraphics[width=\figsizepsferrapp\linewidth,valign=m]{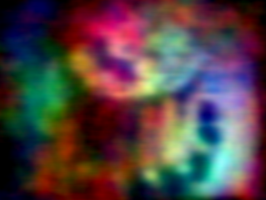}
		&\includegraphics[width=\figsizepsferrapp\linewidth,valign=m]{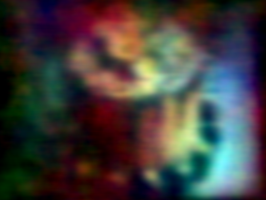}
		\\[\newlinepsfeffapp]
		\includegraphics[width=\figsizepsferrapp\linewidth,valign=m]{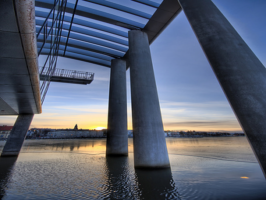}
		&
		\includegraphics[width=\figsizepsferrapp\linewidth,valign=m]{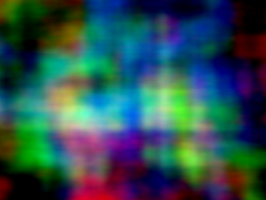}
		&\includegraphics[width=\figsizepsferrapp\linewidth,valign=m]{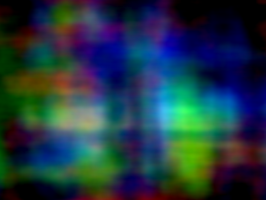}
		&\includegraphics[width=\figsizepsferrapp\linewidth,valign=m]{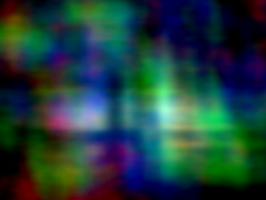}
		&\includegraphics[width=\figsizepsferrapp\linewidth,valign=m]{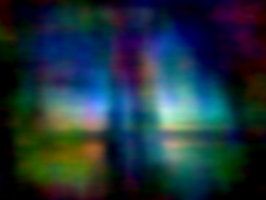}
		&\includegraphics[width=\figsizepsferrapp\linewidth,valign=m]{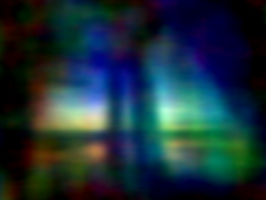}
		&\includegraphics[width=\figsizepsferrapp\linewidth,valign=m]{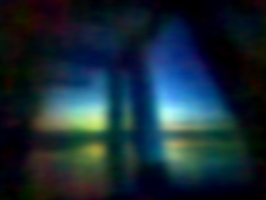}
		\\
	\end{tabular}
	\caption{Example reconstructions with 100 iterations of ADMM as the percentage of correct pixels in the mask pattern increases. Top row shows example PSFs according to the percentage of correct pixels.}
	\label{fig:different_admm}
\end{figure*}

\section{Authentication Confusion Matrices}

\begin{figure*}[t!]
	\centering
	\begin{subfigure}{0.32\linewidth}
		\centering
		\includegraphics[width=0.99\linewidth]{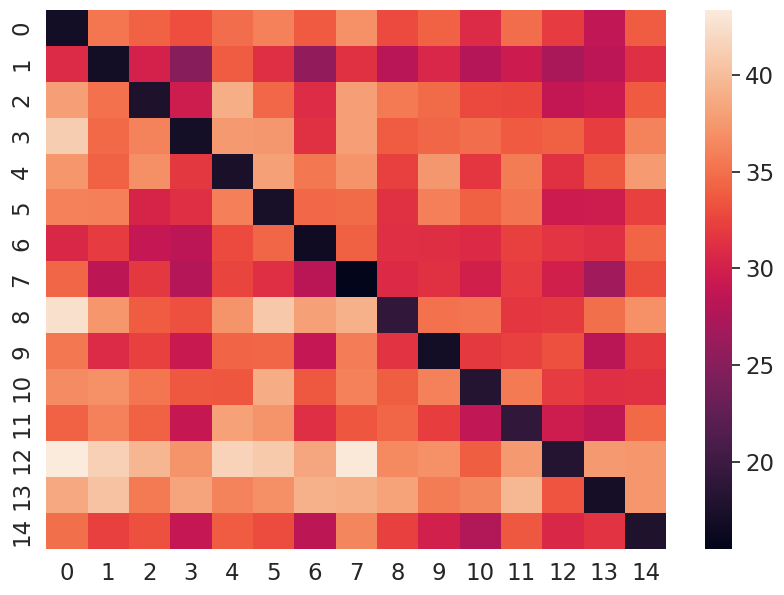} 
		\caption{Data fidelity.}
		\label{fig:auth_admm_data_fid}
	\end{subfigure}
	\begin{subfigure}{0.32\linewidth}
		\centering
		\includegraphics[width=0.99\linewidth]{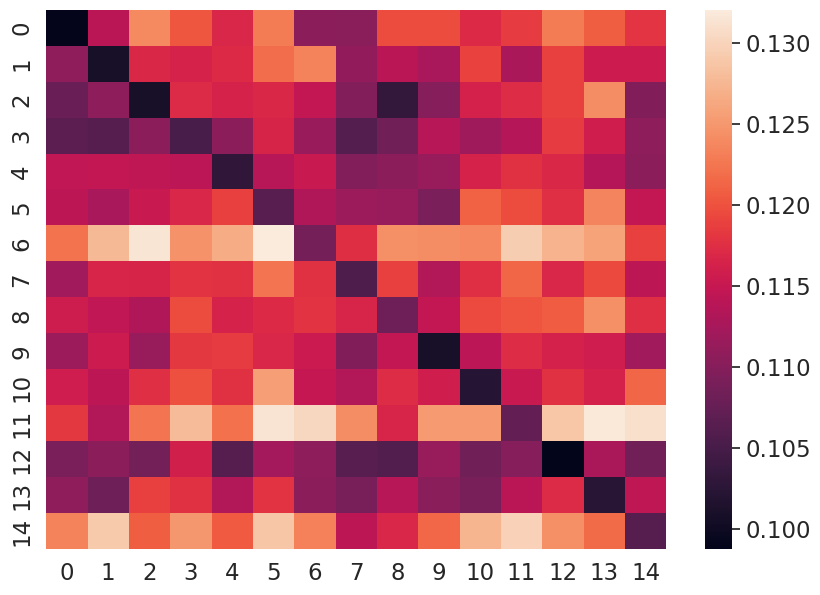} 
		\caption{MSE.}
		\label{fig:auth_admm_mse}
	\end{subfigure}
	\begin{subfigure}{0.32\linewidth}
		\centering
		\includegraphics[width=0.99\linewidth]{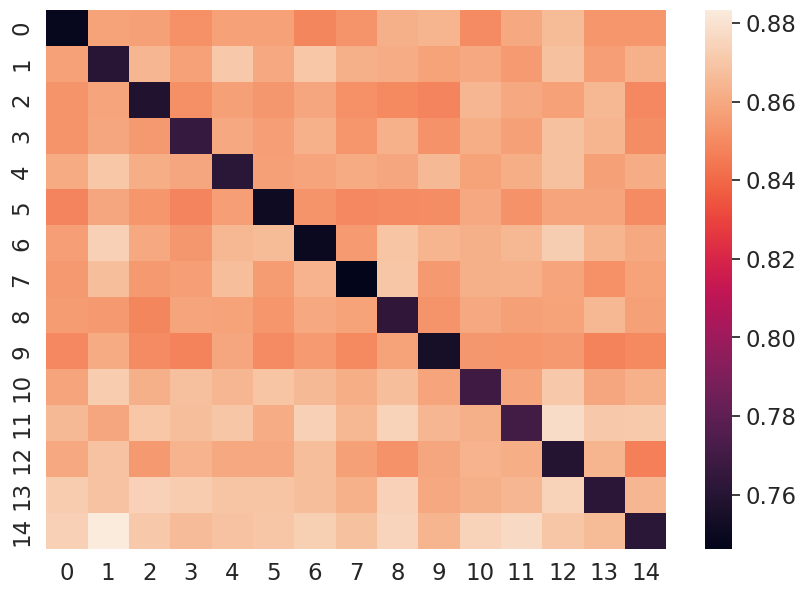} 
		\caption{LPIPS.}
		\label{fig:auth_admm_lpips}
	\end{subfigure}
	\caption{Authentication confusion matrices for different metrics used  when 100 iterations of ADMM is used for image recovery.}
	\label{fig:auth_admm}
\end{figure*}

\begin{figure*}[t!]
	\centering
	\begin{subfigure}{0.32\linewidth}
		\centering
		\includegraphics[width=0.99\linewidth]{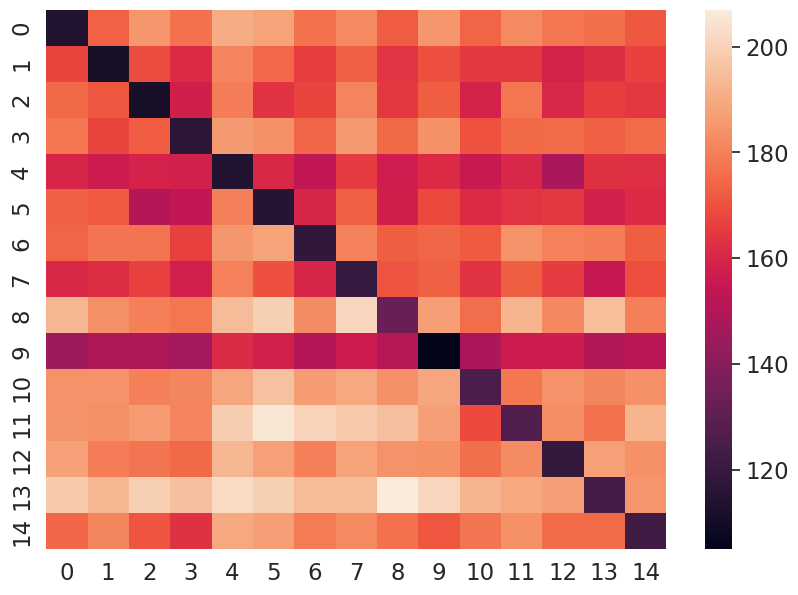} 
		\caption{Data fidelity.}
		\label{fig:auth_learned_data_fid}
	\end{subfigure}
	\begin{subfigure}{0.32\linewidth}
		\centering
		\includegraphics[width=0.99\linewidth]{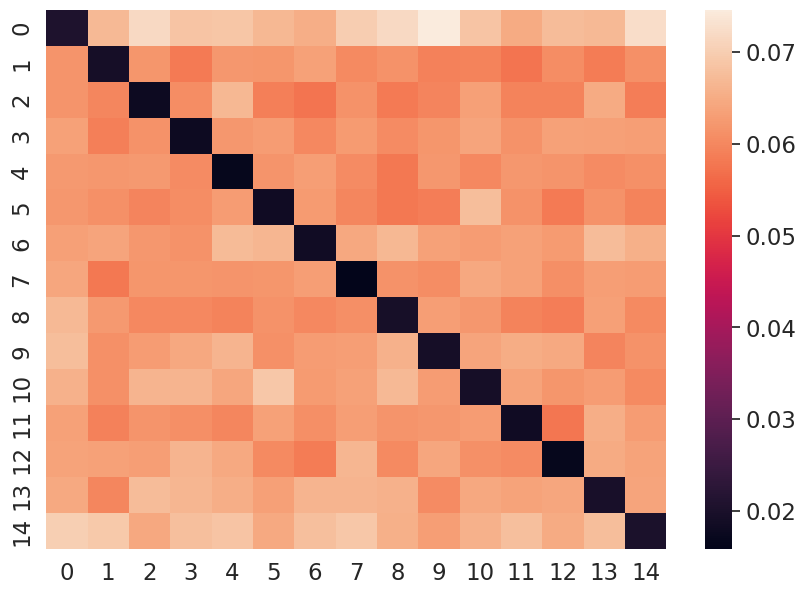} 
		\caption{MSE.}
		\label{fig:auth_learned_mse}
	\end{subfigure}
	\begin{subfigure}{0.32\linewidth}
		\centering
		\includegraphics[width=0.99\linewidth]{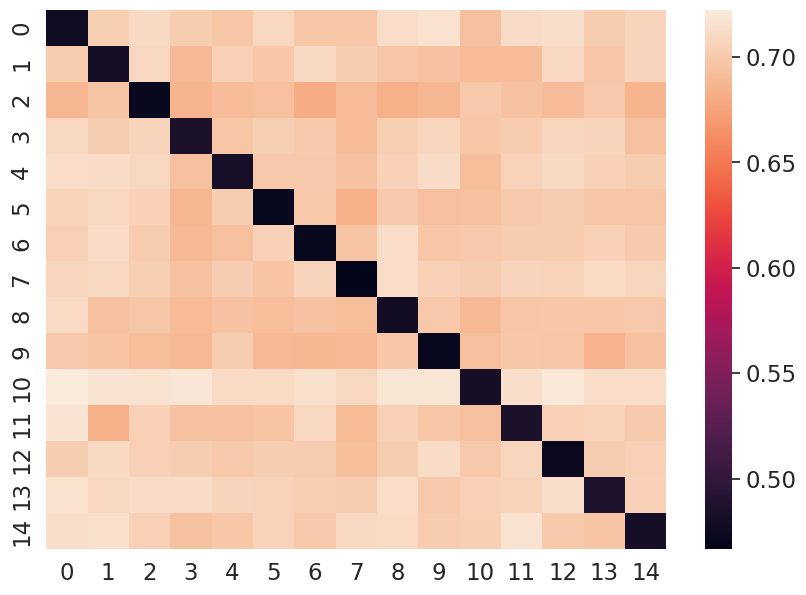} 
		\caption{LPIPS.}
		\label{fig:auth_learned_lpips}
	\end{subfigure}
	\caption{Authentication confusion matrices when a modular learned reconstruction~\cite{Bezzam2024} is used for image recovery.}
	\label{fig:auth_learned}
\end{figure*}

\cref{fig:auth_admm,fig:auth_learned} show the confusion matrices when ADMM and the modular learned reconstruction are used for image recovery, respectively, when performing mask pattern authentication on the \textit{MirFlickr-M} test set.
We can observe that for both approaches,
using the LPIPS metric is better at distinguishing between measurements of different masks.
Using the learned reconstruction yields scores that are further distinguished between masks.

\end{document}